\documentclass[12pt]{amsart} 
\usepackage[margin=1in]{geometry}
\usepackage{graphicx}
\usepackage{multirow}
\usepackage[table,xcdraw]{xcolor}
\usepackage{graphicx,natbib}
\usepackage{amsmath}
\usepackage{amssymb}
\usepackage{epstopdf}
\usepackage{pdfpages}
\usepackage{amsthm}
\usepackage{xcolor}
\usepackage{hyperref}

\newtheorem{theorem}{Theorem}[section]

\newtheorem{proposition}[theorem]{Proposition}

\definecolor{olive}{rgb}{0.5, 0.5, 0.0}

\newcommand{\n}{^{(n)}}
\newcommand{\pms}{{\scriptscriptstyle \pm}}
\newcommand{\cal}{\mathcal}

\newcommand*\X{{\bf X}}
\newcommand*\x{{\bf x}}
\newcommand*\U{{\mathrm U}}
\newcommand*\uu{{\bf u}}

\usepackage{multicol}

\title{
{{\sc Center-Outward Quantiles  \\ and~the ~Measurement of Multivariate~Risk}}}
\date{9 November 2019}

\author[J. \smash{Beirlant}]{Jan Beirlant}
\address[J. Beirlant]{Department of Mathematics, KU Leuven, Belgium}

\email{{jan.beirlant@kuleuven.be}}

\author[S. \smash{Buitendag}]{Sven Buitendag}
\address[S. Buitendag]{Department of Statistics and Actuarial Science, University of Stellenbosch, South Africa}

\email{{svenbuitendag@gmail.com}}

\author[E. \smash{Del Barrio}]{Eustasio del Barrio}
\address[E. Del Barrio]{IMUVA, Universidad de Valladolid, Spain}

\email{{tasio@eio.uva.es}}

\author[M. \smash{Hallin}]{Marc Hallin}
\address[M. Hallin]{ECARES, Universit\'e libre de Bruxelles, Belgium}

\email{{mhallin@alb.ac.be}}

\begin{document}
\maketitle
\begin{abstract} All multivariate extensions of the univariate  theory of risk measurement run into the same fundamental problem of  the absence, in dimension $d>1$, of a canonical ordering of $\mathbb{R}^d$. Based on measure transportation ideas, several attempts have been made recently in the statistical literature to overcome  that conceptual difficulty.  In Hallin~(2017), the concepts of center-outward distribution and quantile functions are developed as generalisations of the classical univariate concepts of  distribution and quantile functions, along with their empirical versions. 
The center-outward distribution function~${\bf F}_{\scriptscriptstyle\pm}$  is a homeomorphic cyclically monotone mapping from  $\mathbb{R}^d\setminus {\bf F}^{-1}_{\scriptscriptstyle\pm}({\bf 0})$ to the open punctured unit ball~$\mathbb{B}_d\setminus\{{\bf 0}\}$, while its empirical counterpart ${\bf F}_{\scriptscriptstyle\pm}\n$  is a cyclically monotone mapping from the sample to a regular grid over $\mathbb{B}_d$. In dimension $d=1$,~${\bf F}_{\scriptscriptstyle\pm}$ reduces to~$2F-1$, while~${\bf F}_{\scriptscriptstyle\pm}\n$ generates the same sigma-field as traditional univariate ranks. The empirical~${\bf F}_{\scriptscriptstyle\pm}\n$, however,  involves a large number of ties, which is impractical in the context of risk measurement. 
 We therefore propose a  class of smooth approximations ${\bf F}_{n,\xi}$ ($\xi$ a smoothness index) of~${\bf F}_{\scriptscriptstyle\pm}\n$ as an alternative to the  interpolation developed in del Barrio et al.~(2018). This approximation allows for the computation of some new empirical risk measures, based either on  the convex potential associated with the proposed transports, or on the volumes of the resulting empirical  quantile regions. We also discuss the role of such transports in the evaluation of the risk associated with multivariate regularly varying distributions. Some simulations and applications to case studies illustrate the value of the  approach. 
\end{abstract}

\section{Introduction}

Any attempt to extend   univariate  risk measurement ideas to a multivariate context  has to face  the preliminary but very fundamental problem of   the absence, in dimension $d>1$, of a canonical ordering of $\mathbb{R}^d$. The absence of such ordering, indeed,  implies the absence of a canonical extension of    fundamental notions as distribution and quantile functions,   expectiles, extreme values, $\ldots$, playing a fundamental role in the definition of such basic tools as (conditional) values at risk, integrated quantile functions, expected shortfalls,  QQ plots, extremograms, Lorenz curves, etc. 

Based on the theory of measure transportation, several ideas have been proposed recently to overcome this absence of a multivariate ordering  in dimension $d$ higher than one. Ekeland et al.~(2012) with their concept of comonotonicity can be considered as  forerunners; Chernozhukov et al.~(2017) with the definition of {\it Monge-Kantorovich depth} and Hallin~(2017) with the  introduction of {\it center-outward distribution and quantile functions} are providing convincing concepts of distribution-specific orderings, with data-driven empirical counterparts and obvious consequences in multivariate risk measurement. The objective of this paper is an investigation  of some of these consequences.
 
\subsection{Center-outward distribution and quantile functions}\label{introsec1}

Center-outward distribution and quantile functions  (Hallin 2017) were introduced in order to circumvent the lack of left-to-right ordering if $\mathbb{R}^d$ in dimension $d$ higher than one. The center-outward distribution   function of a $d$-dimensional Lebesgue-absolutely continuous random vector~$\X$ is defined as the unique gradient of a convex function~${\bf F}_{\scriptscriptstyle\pm}$   pushing the $\X$ distribution  forward to the uniform   $\U_d$ over the unit ball\footnote{By uniform over~$\mathbb{B}_d$ we mean {\it spherical uniform}, i.e.~$\U_d$ is  the product measure of a uniform   over the directions (the unit sphere~$\mathbb{S}_{d-1}$) and a uniform  over the distances to the origin (the unit interval~[0,1]). 
Of course, in case $\bf X$ only takes  positive values,   $\U_d$ is to be replaced with its restriction to the intersection of $\mathbb{B}_d$ and the positive orthant. 
} $\mathbb{B}_d$---as defined in optimal transport theory (for background reading, see, for instance, the monograph by Villani (2009)). The  corresponding quantile function  ${\bf Q}_{\scriptscriptstyle\pm}\!:={\bf F}_{\scriptscriptstyle\pm}^{-1}$ is the inverse of ${\bf F}_{\scriptscriptstyle\pm}\!$ (possibly, a set-valued function). 
 This definition, for $d=1$, yields   the monotone mapping $F_{\scriptscriptstyle\pm} =2F-1$ (with~$F$~the classical univariate distribution function) from the domain of a random variable $X$ to~(-1,1), the open unit ball in $\mathbb{R}$.  
  Clearly,~${F}$ and~${F}_{\scriptscriptstyle\pm}$ carry the same information about the  distribution of~$X$, which they both characterize. 
  
   The {\it center-outward quantile function}~${\bf Q}_{\scriptscriptstyle\pm}$ is mapping the collection~$\{\tau{\mathbb{S}}_{d-1}\vert\ \tau\in [0, 1)\}$ of (hyper)spheres with radii $\tau\in [0,1)$) to the collection~$\{
 {\bf Q}_{\scriptscriptstyle\pm}(\tau \mathbb{S}_{d-1})
\}$ of  {\it center-outward quantile contours}; Hallin~(2017) and, under more general assumptions, Hallin et al.~(2019) show that those contours are continuous, closed, connected, and nested, enclosing {\it quantile regions} 
 with probability content~$\tau$. In dimension $d=1$ and distributions with nonvanishing densities, those quantile regions are nested  interquantile intervals of the form~$[Q((1-\tau)/2), Q((1+\tau)/2)]$,  with $Q:=F^{-1}$ denoting the classical quantile function.
 
An important particular case is provided by the elliptical distributions. Given a location vector ${\boldsymbol \mu} \in \mathbb{R}^d$ and a positive-definite symmetric matrix ${\boldsymbol\Sigma}$, a random vector $\X$ has elliptical distribution if and only if ${\bf Z}:= {\boldsymbol\Sigma}^{-1/2} ({\X}-{\boldsymbol \mu})$ has a spherical distribution, which holds if and only if 
\begin{equation}
{\bf F}_{ell}({\bf Z}) := {{\bf Z} \over \left\| {\bf Z}\right\|} F_R (\left\| {\bf Z}\right\|) \sim \U_d,
\label{Fell}
\end{equation} 
 where $F_R$ denotes the distribution function of $\left\| {\bf Z}\right\|$. 
 Chernozhukov et al.~(2017) show that~${\bf F}_{ell}$ corresponds to the center-outward distribution function  ${\bf F}_{\scriptscriptstyle\pm}$ of ${\bf Z}$.  
For the quantile function we then have, denoting by ${\bf U}$  a   random vector with distribution $\U_d$, 
 \begin{equation}
{\bf Q}_{\scriptscriptstyle\pm}({\bf U}) ={\bf Q}_{ell}({\bf U}) := {{\bf U} \over \left\| {\bf U}\right\|} Q_R (\left\| {\bf U}\right\|) =_d  {\bf Z},
\label{Qell}
\end{equation} 
 where $Q_R:=F_R^{-1}$ denotes the quantile function of $\left\| {\bf Z}\right\|$. 
 \\

Turning to sampling values, denote by $\X\n:=\big(\X_1,\ldots, \X_n\big)$ an i.i.d.~sample from a population with center-outward distribution function ${\bf F}_{\scriptscriptstyle\pm}$. 
The empirical counterpart~${\bf F}^{(n)}_{\scriptscriptstyle\pm}$ 
  to ${\bf F}_{\scriptscriptstyle\pm}$
  ---more precisely, its restriction ${\bf F}\n_{\scriptscriptstyle\pm}(\X_1),\ldots,{\bf F}\n_{\scriptscriptstyle\pm}(\X_n)$ to  the sample values---  can be obtained as a cyclically monotone (discrete) mapping from the random sample $\X_1,\ldots, \X_n$ to some  ``uniform" grid~${\mathfrak u}\n=\{\mathbf{u}_1,\ldots,\mathbf{u}_n\}$ over $\mathbb{B}_d$.   Writing ${\mathrm U}_{d}\n$ for the empirical measure on the $n$ gridpoints, the only technical requirement is that the grid~${\mathfrak u}\n$ is such that 
\begin{equation}
\label{ConsistencyCondition}
{\mathrm U}_{d}\n\longrightarrow_w {\mathrm U}_{d}\qquad\text{as $n\to\infty$,}
\end{equation}
where $\longrightarrow_w$ denotes weak convergence of probability measures. 

In order to obtain such a grid,  Hallin (2017)  factorizes the sample size into  $n~\!\!=\!~\!n_Rn_S~\!+~\!n_0$ with~$0\leq n_0<\min(n_R, n_S)$: $n_S$ unit vectors ${\bf s}_1,\ldots,{\bf s}_{n_S}$ (that is, $n_S$ points on the unit sphere) are chosen---for instance, randomly  generated from the uniform distribution over~$\mathbb{S}_{d-1}$---each of them scaled up to the $n_R$ radii $r\in \{ \frac 1{n_R+1},\ldots,\frac{n_R}{n_R+1}\}\vspace{0.8mm}$. Along with~$n_0$ copies of the origin, the  resulting $n_Rn_S$ points  constitute a grid ${\mathfrak u}\n$ that satisfies~\eqref{ConsistencyCondition} as both $n_R$ and $n_S$ tend to infinity (which implies $n_0/n\to 0$).   

The values ${\bf F}\n_{\scriptscriptstyle\pm}(\X_1),\ldots,{\bf F}\n_{\scriptscriptstyle\pm}(\X_n)$ of the empirical center-outward  distribution function are constructed as the solution of the discrete optimal  transport  problem pushing the uniform distribution over the sample $\X\n$ forward to the uniform over the grid ${\mathfrak u}\n$. More precisely, they satisfy
\begin{equation}\label{optass1} 
\sum_{i=1}^n\big\Vert {\bf X}_i - {\bf F} _{\pms}\n({\bf X}_i)
\big\Vert ^2
=
 \min_{T\in{\cal T}} \sum_{i=1}^n\big\Vert {\bf X}_{i} - T({\bf X}_i)
\big\Vert ^2\vspace{-1mm}
\end{equation}
where the minimum is taken over the set $\cal T$  of all possible bijective mappings $T$ between~$\X\n$ and  the   grid ${\mathfrak u}\n$ or, equivalently, 
\begin{equation}\label{optass2} 
\sum_{i=1}^n\big\Vert {\bf X}_i - {\bf F} _{\pms}\n({\bf X}_i)
\big\Vert^2 
=
 \min_\pi \sum_{i=1}^n\big\Vert {\bf X}_{\pi (i)} - {\bf F} _{\pms}\n({\bf X}_i)
\big\Vert ^2\vspace{-1mm}
\end{equation}
where the set $\{{\bf F} _{\pms}\n({\bf X}_i)\vert\ i=1,\ldots ,n\}$ coincides with the the set of~$n$ gridpoints in ${\mathfrak u}\n$ and 
$\pi$ ranges over the $n!$ possible permutations of $\{1,2,\ldots ,n\}$.

Classical results (see  McCann~(1995)) then show that    \eqref{optass1}-\eqref{optass2} is satisfied iff     {\it cyclical monotonicity}\footnote{Recall that  a subset  $S:=\{(\mathbf{x}_1,{\bf y}_1),\ldots,(\mathbf{x}_n,{\bf y}_n)\}$  of $\mathbb{R}^d\times \mathbb{R}^d$  is said to be {\em cyclically monotone} if, for any finite collection 
 $\{({\bf x}_{i_1}, {\bf y}_{i_1}),\ldots ,({\bf x}_{i_k}, {\bf y}_{i_k})\}\subseteq S$, 
 $
 \langle {\bf y}_{i_1},\ {\bf x}_{i_2}-{\bf x}_{i_1}\rangle + \langle {\bf y}_{i_2},\ {\bf x}_{i_3}-{\bf x}_{i_2}\rangle +\ldots + \langle {\bf y}_{i_k},\ {\bf x}_{i_1}-{\bf x}_{i_k}\rangle  \leq 0.
$ 
} 
  holds for the $n$-tuple 
\begin{equation}\label{optcouples} \big\{\big({\bf X}_1, {\bf F} _{\pms}\n({\bf X}_1)\big),\ldots , \big({\bf X}_n, {\bf F} _{\pms}\n({\bf X}_n)\big)\big\}.
\end{equation}
It is well known that the subdifferential of a convex function $\Psi$ from $\mathbb{R}^d$ to $\mathbb{R}$ enjoys cyclical monotonicity, while the converse is also true as shown by Rockafellar (1966): any cyclically monotone subset of $\mathbb{R}^d\times \mathbb{R}^d$ is contained in  the subdifferential of some convex function---the interpolations we are describing below are constructive applications of that fact.

This ${\mathfrak u}\n$-based construction of  ${\bf F}\n_{\scriptscriptstyle\pm}$, however, presents two major drawbacks: the resulting quantile contours indeed 
\begin{enumerate}
\item[(a)]  are discrete collections of $n_S$ observations whereas graphical quantile representations and our risk analysis objectives require connected continuous surfaces enclosing compact  nested regions, and 
\item[(b)]    yield an unpleasantly high number $n_S$ of tied observations (the $n_S$ observations in a given quantile contour).  
\end{enumerate}

A remedy to (a) is smooth interpolation, under cyclical monotonicity constraints, of the $n$-tuples $({\bf X}_1, {\bf F} _{\pms}\n({\bf X}_1)),\ldots,({\bf X}_n, {\bf F} _{\pms}\n({\bf X}_n))$. This program was carried out in del Barrio et al.~(2018),   based on the concept of {\it Moreau envelopes}, providing smooth extensions $\widetilde{\bf F} _{\pms}\n$   of ${\bf F} _{\pms}\n$ (mapping $\mathbb{R}^d$ to $\mathbb{B}_d$) and $\widetilde{\bf Q} _{\pms}\n$ of  its inverse ${\bf Q} _{\pms}\n$ (mapping $\mathbb{B}_d$  to $\mathbb{R}^d$). From the point of view of risk measurement, those interpolations, unfortunately, do not allow for an easy calculation of the volumes of the corresponding quantile regions to be used in Section~3.  An alternative solution---not a strict interpolation, but an asymptotically equivalent  approximation thereof---allowing for an integral representation of those volumes   is provided in Section~2. 

In order to palliate (b), we propose to compute ${\bf F} _{\pms}\n$ as previously, albeit from another type of random grid. Those grids---denoted as~${\mathfrak w}\n$---are obtained  by generating~$n$ points~${\bf s}_1,\ldots,{\bf s}_n$ from the uniform distribution over~$\mathbb{S}_{d-1}$, then randomly rescaling each of the corresponding unit vectors   to one of  the $n$ radii~$r\in \{ \frac 1{n+1},\ldots,\frac{n}{n+1}\}\vspace{0mm}$; such  grids obviously satisfy\eqref{ConsistencyCondition} as~$n\to\infty$. The  empirical center-outward distribution function resulting from pushing the empirical distribution forward to the uniform over~${\mathfrak w}\n$ now has~$n$ distinct contours and no ties: each contour and each sign curve indeed, with probability one, consists of one single observation. After due interpolation or approximation,~$n$ quantile contours (orders $\frac{1}{n+1},\ldots, \frac{n}{n+1}$) are obtained (in the case of an interpolation, each of them  running through  one single observation);  contours of arbitrary intermediate orders are available as well. 

%

Whether resulting from interpolation or from the approximations considered in Section~2, the empirical and  smoothed empirical center-outward distribution functions based on grids of the ${\mathfrak w}\n$ type enjoy  the same asymptotic properties\footnote{A Glivenko-Cantelli property was established  (without any moment conditions) for ${\bf F}^{(n)}_{\scriptscriptstyle\pm}$   by Hallin~(2017) 
 and extended to its interpolated version    $\widetilde{\bf F}_n$ 
  by del~Barrio et al.~(2018). 
} 
as those based on grids of the ${\mathfrak u}\n$ type. For finite $n$, however, the random construction of   ${\mathfrak w}\n$   may lead to poor results due to a very bad matching of the random ${\mathfrak w}\n$ with the actual sample. In practice, this is easily taken care of by  {\it (i)} independently generating $m$  grids ${\mathfrak w}\n_1, \ldots, {\mathfrak w}\n_m$, {\it (ii)} computing the corresponding~${\bf F} _{\pms; k}\n({\bf X}_i)$ ($k=1,\ldots,m$,  $i=1,\ldots, n$), {\it (iii)} averaging, for each ${\bf X}_i$, those $m$ ${\bf F} _{\pms; k}\n({\bf X}_i)$ values  into~${\overline{\bf F}} _{\pms}^{(n)m}({\bf X}_i):=m^{-1}\sum_{k=1}^m {\bf F} _{\pms; k}\n({\bf X}_i),$ 
 and {\it (iv)} performing, as explained in del Barrio etal.~(2018), a smooth interpolation   under cyclical monotonicity constraints  ($\widetilde{\bf F}_n$, say)  of the $n$-tuple~$({\bf X}_1, {\overline{\bf F}} _{\pms}^{(n)m}({\bf X}_1)),\ldots,({\bf X}_n, {\overline{\bf F}} _{\pms}^{(n)m}({\bf X}_n))$. 
 
 For the sake of simplicity and without any loss of generality, we keep the notation~${{\bf F}} _{\pms}\n$ for~${\overline{\bf F}} _{\pms}^{(n)m}$, with inverse~${\bf Q}_{\pms}^{(n)}$, as if $m$ were equal to one, and even throughout simplify it to ${\bf F}_{n}$ and ${\bf Q}_{n}$:
  all theoretical results below indeed remain valid  for~${\overline{\bf F}} _{\pms}^{(n)m}$, irrespective of $m$, iff they hold for~${{\bf F}} _n:={{\bf F}} _{\pms}\n$. As for the numerical results, they all were obtained for~$m=10$.

 \subsection{Risk measurement} 
 The literature on  multivariate risk measurement is much less abundant than its univariate counterpart.  From a recent review by Charpentier (2018) it appears that the only  fundamental contribution available on this topic is the one developed in   the pioneering work  by 
 Ekeland et al.~(2012) with their theory of comonotonic  measures of multivariate risk. 
 
 All univariate risk measures are directly or indirectly related to quantiles, hence to the left-to-right ordering of $\mathbb R$. In the univariate case the empirical quantile function immediately allows for constructing nonparametric estimates of  such risk measures   as Value-at-Risk and expected shortfall. 
Although  Ekeland et al. (2012) clearly showed how the optimal transport theory offers a well-founded path to  multivariate  generalizations of  quantiles,  practical implementation has not taken up. Empirical multivariate risk measurement still mostly relies on copula modelling followed by the computation of some ad hoc  risk functionals. Our objective, in this paper, is to provide novel ways  of measuring multivariate risk, whether nonparametrically or by evaluating tail heaviness   in   multivariate regularly varying models. Our approach, in the spirit of  Ekeland et al. (2012), is based on (interpolations or adequate approximations of) the empirical  center-outward quantile function ${\bf Q} _{n}$ and, more particularly, the scalar products~$\langle {\bf u}_i, {{\bf Q}} _{n} ({\bf u}_i) \rangle,\; i=1,\ldots ,n$. 

\subsection{Outline of the paper} 

 In Section 2, we define and study the main theoretical  tool to be used in our approach to multivariate risk measurement. Essentially, we introduce a smooth cyclically monotone approximation $\widehat{\bf Q}_{n,\xi}$ ($\xi$ a smoothness parameter) of the (discretely defined) empirical center-outward quantile function ${\bf Q}_{n}$ considered in Hallin~(2017) and del Barrio et al.~(2018). Theorem~\ref{consistency} establishes the a.s.~consistency, uniformly over compacts, of this approximation.

  In Section 3, we consider empirical measures of multivariate risk   based on $\widehat{\bf Q}_{n,\xi}$,  the corresponding  potential 
$\Psi_{n,\xi}$, and the volumes of the resulting quantile regions. The potential then leads to generalizations of risk measures based on the (univariate) integrated quantile functions, as discussed in detail in Gushin and Borzykh (2018). On the other hand, using   ${\bf Q}_{n,\xi}= \nabla \Psi_{n,\xi}$ leads to empirical  risk measurements of the  maximal correlation type   $\mathrm{E}\left(\langle {\bf U},\nabla \Psi ({\bf U})\rangle \right)$ as discussed in Ekeland et al.~(2012). 
 Section~4 shows how to compute 
 the volumes of the empirical contours characterized by  $\widehat{\bf Q}_{n,\xi}$. Those volumes can be used in an extension to the multivariate context  of traditional univariate QQ plots---a widespread  tool in the analysis of univariate risk.   Finally, Section~5 discusses some applications of our measure transportation approach  to risk measurement for multivariate regularly varying distributions.  Simulation-based illustrations of the proposed concepts are provided throughout Sections 2-5.  
Section~6 concludes   with two real-life  applications from finance and  insurance, demonstrating  the practical value  of the proposed methods.

\section{Cyclically monotone interpolation   and cyclically monotone approximation}

\subsection{A smooth cyclically monotone approximation of empirical quantiles}
In this section, we construct  smooth   approximations $\widehat{\bf Q}_{n,\xi}$ of   ${\bf Q}_n$ as an alternative to the smooth interpolations $\widetilde{\bf Q}_n$ proposed in del Barrio et al.~(2018). Those approximations depend   on a smoothing parameter $\xi$, and allow  for an integral representation of the volumes of the corresponding quantile regions. Along with $\widehat{\bf Q}_{n,\xi}$, we  also construct (up to an additive constant) the convex  potential $\Psi_{n,\xi}$ such that $\widehat{\bf Q}_{n,\xi}=\nabla \Psi_{n,\xi}$. 
 While these approximations~$\widehat{\bf Q}_{n,\xi}$ are not proper extensions of ${\bf Q}_n$ in the sense that they do not map the   gridpoints they are built from to the sample, we show that,  for suitable choices of the smoothing parameter $\xi$, they nevertheless provide strongly consistent estimators of the actual center-outward quantile function ${\bf Q}_{\pms}$.

The smooth cyclically monotone interpolation,  
 $\widetilde{\bf Q}_n$, say, 
 proposed in del Barrio et al.~(2018), on the contrary, constitutes a proper extensions of ${\bf Q}_n$ and is obtained    as follows.  Throughout, let  $\X\n=(\X_1,\ldots,\X_n)$ denote a sample from a population with center-outward and quantile functions ${\bf F}_\pms$ and ${\bf Q}_\pms$, respectively, and  consider an arbitrary sequence ${\mathfrak u}\n:=\{{\bf u}_1,\ldots,{\bf u}_n\}$    of grids satisfying \eqref{ConsistencyCondition}; in this section, the $\X_i$'s safely can be treated as constants. 
 
 By definition, the empirical quantile function ${\bf Q}_n$ obtained from transporting  ${\mathfrak u}\n$ to~$\X\n$ is such that~${\bf Q}_n({\bf u}_i)=\X_i$, $i=1,\ldots,n$ and the set $\{({\bf u}_i,\X_i):\, i=1,\ldots,n\}$ is cyclically monotone. 
{ Hence, almost surely (see Proposition 2.1 in del Barrio et al.~(2018)),} there exist constants $\lambda_1,\ldots ,\lambda_n\in \mathbb{R}$ such that the linear functions mapping~$\bf u$ to~$\psi_i({\bf u}) :=\langle {\bf u},\X_i\rangle - \lambda_i $
satisfy 
\begin{align}
\psi_i\left({\bf u}_i \right) -\psi_j\left({\bf u}_i \right) > 0\hspace{2mm} \text{ for all } \hspace{2mm} i\not=j \in \{1,2,\dots,n\}. \label{psi condition}
\end{align} 
 The map $\Psi_n$ defined as
 $  {\bf u} \mapsto \Psi_n ({\bf u}) := \max_{i=1,\dots,n}
  \psi_i({\bf u})
  $  
  is thus a convex map and the open  sets $C_i=\left\{ {\bf u} \in \mathbb{S}_d \, | \, \psi_i\left({\bf u} \right)
> \max_{j \neq i}\psi_j\left({\bf u} \right)   \right\}$ are convex sets. Moreover, $\Psi_n $ is differentiable on $C_i$, with gradient $\nabla \Psi_n ({\bf u})= \X_i$ for ${\bf u}\in C_i$, and \eqref{psi condition} entails~$\nabla \Psi_n ({\bf u}_i)=~\!\X_i$.\linebreak  
Hence,~$\nabla \Psi_n$ is a piecewise constant extension of ${\bf Q}_n$ to $\cup_{i=1}^n C_i$, for which we keep the same notation, namely,~${\bf Q}_n({\bf u}):= \nabla \Psi_n ({\bf u})$.
\vspace{1mm}

Now,  condition \eqref{psi condition} is equivalent to 
$$\langle {\bf u}_i, \x_{i} - \x_{j} \rangle  > \lambda_i - \lambda_j\quad\text{ for all $i\not=j \in \{1,2,\dots,n\}$. }$$
It   follows that the constants $\lambda_1, \lambda_2, \dots, \lambda_n$ are  solutions to the  linear program maximizing~$\delta$ subject to
\begin{align*}
\left[ \begin{array}{c c c c c c} 	1 & -1 & 0 & \dots & 0 & 1 \\
													1 & 0	 & -1 &  \dots & 0 & 1 \\
													\vdots & \vdots & \vdots  & \ddots & \vdots & \vdots \\
													1 & 0 & 0 & \dots & -1 & 1 \\
													-1 & 1 & 0 & \dots & 0 & 1 \\ 
													0 & 1 & -1 & \dots & 0 & 1 \\ 
													\vdots & \vdots & \vdots  & \ddots & \vdots & \vdots \\
													0 & 1 & 0 &  \dots & -1 & 1 \\ 
													-1 & 0 & 1 & \dots & 0 & 1 \\ 
													0 & -1 & 1 & \dots & 0 & 1 \\ 
													\vdots & \vdots  & \vdots & \ddots & \vdots & \vdots \\
													-1 & 0 & 0  & \dots & 1 & 1 \\
													0 & -1 & 0 & \dots & 1 & 1 \\
													0 & 0 & -1 &  \dots & 1 & 1 \\
													\vdots & \vdots & \vdots  & \ddots & \vdots & \vdots \\
													0 & 0 & 0 &  \dots & 1 & 1 \end{array} \right]
	\, \left[ \begin{array}{c} \lambda_1 \\ \lambda_2 \\ \vdots \\ \lambda_n \\\ \delta \end{array} \right]
	 \leq \left[ \begin{array}{c} 	\langle {\bf u}_1 , \X_{1} - \X_{2} \rangle \\
							\langle {\bf u}_1 , \X_{1} - \X_{3} \rangle \\
							\vdots  \\
							\langle {\bf u}_1 , \X_{1} - \X_{n} \rangle \\
							\langle {\bf u}_2 , \X_{2} - \X_{1} \rangle \\ 
							\langle {\bf u}_2 , \X_{2} - \X_{3} \rangle \\ 
							\vdots \\
							\langle {\bf u}_2 , \X_{2} - \X_{n} \rangle \\ 
							\langle {\bf u}_3 , \X_{3} - \X_{1} \rangle \\ 
							\langle {\bf u}_3 , \X_{3} - \X_{2} \rangle \\ 
							\vdots \\
							\langle {\bf u}_n , \X_{n} - \X_{1} \rangle \\
							\langle {\bf u}_n , \X_{n} - \X_{2} \rangle \\
							\langle {\bf u}_n , \X_{n} - \X_{3} \rangle \\
							\vdots \\
							\langle {\bf u}_n , \X_{n} - \X_{n-1} \rangle \end{array} \right].
\end{align*}
The solution $\delta$ of that program satisfies 
\begin{align*}
\delta =  \min_{i=1,\dots,n} \Big\{ \psi_i\left( {\bf u}_i \right) - \max_{j\not=i} \psi_j \left( {\bf u}_i  \right) \Big\} > 0,
\end{align*}
hence  is the minimum difference between $\psi_i\left({\bf u}_i  \right)$ and $\psi_j \left( {\bf u}_i  \right)$, $i\not=j$.

 Let $\Psi_{n,\epsilon} ({\bf u}) := \inf_{{\bf v} \in \mathbb{S}_d}\left\{ \Psi_n ({\bf v})+{1 \over 2 \epsilon}
\left\| {\bf u}-{\bf v}  \right\|^2 
 \right\}$, $\epsilon >0$. The {\it Moreau envelope} on  $\Psi_n$ is defined as 
$
 \widetilde{\bf Q}_{n,\epsilon}({\bf u}):= \nabla 
  \Psi_{n,\epsilon} ({\bf u})$ 
and satisfies 
 \begin{align}
 \widetilde{\bf Q}_{n,\epsilon}({\bf u})=
 \sum_{i=1}^n w_{i,\epsilon}  ({\bf u})\X_i\label{interpolation}
 \end{align}
 where the weights $w_{i,\epsilon}, \; i=1,\ldots,n $ are a solution to the maximization problem
 \begin{align*}
 \max_{w_1,\ldots,w_n} \Big( \sum_{i=1}^n w_{i}\psi_i({\bf u})
 -{\epsilon \over 2}
 \Big\| \sum_{i=1}^n w_{i}\X_i \Big\|^2 \Big)
\  \mbox{ with }\  0 \leq w_i \leq 1 \mbox{ and }  \sum_{i=1}^n w_{i}=1.
 \end{align*}
This problem  can be solved by using a gradient descent algorithm.
For $\epsilon$ sufficiently small (viz., $0<\epsilon \leq \epsilon_0$, where $\epsilon_0$, a data-driven quantity, is characterized in Corollary~2.4 in del Barrio et al.~(2018)),~$\widetilde{\bf Q}_{n,\epsilon}$ then provides (see del Barrio et al.~(2018) for a proof) a continuous, cyclically monotone interpolation of $({\bf u}_i,\X_i), \; i=1,\ldots,n$.  From smoothness considerations the choice in that work was $\widetilde{\bf Q}_n:=\widetilde{\bf Q}_{n,\epsilon_0}$.

As explained in Section~\ref{introsec1}, $\widetilde{\bf Q}_{n}$ unfortunately yields too many ties for our needs in risk measurement.  We therefore  consider a class of alternative weights involving  continuous transformations of~$\psi_1,\ldots,\psi_n$. Namely, let 
\begin{align}
\widehat{\bf Q}_{n,\xi}({ \bf u}) & := \frac{1}{\xi} \, \nabla  \Big(\log  \sum_{i=1}^n e^{\xi \psi_i({\bf u})} \Big) 
	 =\frac{ \sum_{i=1}^n e^{\xi \, \psi_i({\bf u})} \,   \X_i }{\sum_{i=1}^n e^{\xi \, \psi_i({\bf u})}} 
 	 =: \sum_{i=1}^n w_{i,\xi}({\bf u}) \,   \X_i, 
 	\label{sven}
\end{align}
where $\xi>0$ is a smoothing parameter. 
The corresponding potential function is 
\begin{equation}
\Psi_{n,\xi} ({\bf u})= \frac{1}{\xi} \, \log  \sum_{i=1}^n e^{\xi \psi_i({\bf u})}
\label{TPsi_n}
\end{equation}
which, being the logarithm of a sum of exponentials of convex functions, is convex (see Example 3.14, p.~87 of Boyd and Vandenberghe (2004)).  
Hence, $\widehat{\bf Q}_{n,\xi}$ is a smooth cyclically monotone map 
and the corresponding quantile contours    are closed, connected, and nested.  Note also that letting $\xi \to \infty$ brings, for any fixed $n$, this smoothed empirical quantile function arbitrarily close to the non-smooth  piecewise constant one ${\bf Q}_n$   (in the sense that $\lim_{\xi\to\infty}\widehat{\bf Q}_{n,\xi}({\bf u})={\bf Q}_{n}({\bf u})$ for every ${\bf u}\in\cup_{i=1}^n C_i$,  hence for almost every ${\bf u}$) while, as~$\xi \to 0$ (fixed $n$), $\widehat{\bf Q}_{n,\xi}$ approaches the improper (constant) quantile function mapping~$\uu$ to $\overline{\bf X}_n:= \frac{1}{n}\sum_{i=1}^n {\bf X}_i$,  the ultimate smooth version of $\widehat{\bf Q}_{n,\xi}$. 

A closer look at equations \eqref{interpolation} and \eqref{sven} reveals that  the smooth  interpolations $\widetilde{\bf Q}_{n,\epsilon}$    and the smooth approximations $\widehat{\bf Q}_{n,\xi}$ proposed here share some common features. Both are cyclically monotone maps with values in the convex hull of the observed $\X_i$'s.
But there are also some important differences. For $\epsilon\in (0,\epsilon_0]$, 
 $\widetilde{\bf Q}_{n,\epsilon}$ is an extension of~$\mathbf{Q}_\pm\n$ since~$w_{i,\epsilon}(\mathbf{u}_i)=1$ and 
$w_{j,\epsilon}(\mathbf{u}_i)=0$ for~$j\ne i$. For  $\widehat{\bf Q}_{n,\xi}$, on the contrary,  we have~$w_{i,\xi}(\mathbf{u}_i)\!\to\!~\!\!1$
and~$w_{j,\epsilon}(\mathbf{u}_i) \to 0$ as $\xi\to\infty$ but also $w_{i,\xi}(\mathbf{u})\in(0,1)$ for all $i$ and~$\mathbf{u}$. This means that, 
in general, $\widehat{\bf Q}_{n,\xi}(\mathbf{u}_i)\ne \X_i$, so that~$\widehat{\bf Q}_{n,\xi}$ is not an extension of~$\mathbf{Q}_\pm\n$ but, rather, an approximation to $\mathbf{Q}_n$. 

Nevertheless, Theorem~\ref{consistency} below shows that, with a suitable choice of a sequence~$\xi_n$ of smoothing parameters,  $\widehat{\bf Q}_{n,\xi_n}$ remains a consistent estimator of the actual center-outward quantile function ${\bf Q}_\pm$. Additionally, as we will see in the subsequent sections, 
$\widehat{\bf Q}_{n,\xi}$, as a quantile function, is particularly well suited for the computation of   empirical risk measures based on the volume of quantile regions.

For convenience, 
some key properties of $\Psi_{n,\xi}$ are collected in the next proposition. Potentials being defined up to an arbitrary additive constant, let us impose, without any loss of generality,  $\Psi_{n}({\bf 0})=0$  (recalling \eqref{psi condition},  observe that the sets $C_i$ introduced after \eqref{psi condition} remain unchanged if the constants $\lambda_j$ are replaced by $\lambda_j+C$, $j=1,\ldots,n$).
 
{
\begin{proposition}\label{propTech} Denote by $\Psi_n$ the potential associated   with  ${\bf Q}_{n}$, by  
$\Psi_{n,\xi}$ the potential~\eqref{TPsi_n} 
 associated    with $\widehat{\bf Q}_{n,\xi}$. Then, for all $n$ and  $\xi >0$, 
\begin{enumerate}
\item[\it (i)] ${\bf u}\mapsto\Psi_{n,\xi}({\bf u})$ is convex  and differentiable on the open unit ball $\mathbb{B}_d$.
\item[\it (ii)] $\displaystyle{   \Psi _{n} ({\bf u})\leq  \Psi_{n,\xi} ({\bf u}),  {\bf u}\in{\mathbb{B}_d} \mbox{ and } \sup_{{\bf u}\in{\mathbb{B}_d}}\left\vert \Psi_{n,\xi} ({\bf u}) - \Psi _{n} ({\bf u})\right\vert \leq \frac{\log n}{\xi}}$.
\end{enumerate}
\end{proposition}
}

\noindent\textbf{Proof.}   Part~\textit{(i)} of the proposition readily follows from \eqref{sven}, \eqref{TPsi_n},  and subsequent comments. 
To prove part \textit{(ii)}    first note that, 
for every $\xi>0$, $e^{\xi\Psi_n(\mathbf{u})}\leq \sum_{i=1}^n e^{\xi
\psi_i(\mathbf{u})}$, which implies $\Psi _{n} ({\bf u})\leq  \Psi_{n,\xi} ({\bf u})$. On the other hand, $\psi_i(\mathbf{u})\leq 
\Psi_n(\mathbf{u})$ for $i=1,\ldots,n$. Hence, 
$$\sum_{i=1}^n e^{\xi \psi_i(\mathbf{u})}\leq n e^{\xi \Psi_n(\mathbf{u})};$$
it follows that $ \Psi_{n,\xi}(\mathbf{u})-\Psi_{n}(\mathbf{u})
\leq \frac{\log n}{\xi}$ for all $\mathbf{u}$.
\qed\medskip

Under mild regularity assumptions on the distribution of the $\mathbf{X}_i$'s, the center-outward distribution function 
$\mathbf{F}_{\scriptscriptstyle\pm}$ and its inverse, the quantile function $\mathbf{Q}_{\pm}$, are continuous. More precisely, $\mathbf{F}_{\pm}$ and $\mathbf{Q}_{\pm}$ are homeomorphisms between ${\mathbb{B}_d}\setminus\{{\bf 0}\}$ and $\mathbb{R}^d\setminus \mathbf{Q}_{\pm}(\{{\bf 0}\})$, see del Barrio et al.~(2018). A sufficient condition for this is (Figalli~(2018); see del Barrio et al.~(2019) for an extension) that the distribu\-tion~$\mathrm{P}_{\X}$ of $\mathbf{X}_i$ belongs to the class $\mathcal{P}_d$
of probabilities with   Lebesgue density $f$   such that for every~$D>0$ there exist strictly positive constants~$\lambda_{f,D}$ and $\Lambda_{f,D}$ 
such that
 $\lambda_{f,D}<f(\mathbf{x})<\Lambda_{f,D}$
for all $\|\mathbf{x}\|\leq D$. 
Under such assumptions and a proper choice of a sequence~$\xi_n $ of smoothing parameters,   the smoothed empirical potentials $\Psi_{n,\xi_n}$ and the corresponding quantiles $\widetilde{\bf Q}_{n,\xi_n}$ are consistent estimators of their 
population counterparts. This is the content of Theorem~\ref{consistency}, part~{\it (ii)} of which is the quantile version of the Glivenko-Cantelli theorem for center-outward distribution functions. The main difference is that center-outward quantile functions, contrary to the center-outward distribution functions, need not have a bounded range; 
uniform convergence, therefore, is limited to strict subsets of the unit ball---a  limitation that also shows up in the case of classical univariate   quantiles, where uniform convergence typically holds on intervals of type $[u,v]\subset (0,1)$.

{\begin{theorem}\label{consistency} Let the $\X_i$'s be i.i.d.~with distribution   $\mathrm{P}_{\X}\in \mathcal{P}_d$. If  the sequence $\xi_n$ is such that~$
\lim_{n\to\infty}  {\xi_n} /{\log n}=\infty$, 
then, 
\begin{enumerate}
\item[\it (i)] $\displaystyle{  \sup_{{\bf u}\in{\mathbb{B}_d}}\left\vert \Psi_{n,\xi_n} ({\bf u}) - \Psi _{n} ({\bf u})\right\vert   = o(1)}$ as $n\to\infty$;
\item[\it (ii)] for every compact $K\subset \mathbb{B}_d\setminus \{\mathbf{0} \}$,  
$$\sup_{\mathbf{u}\in K} \| \widehat{\mathbf{Q}}_{n,\xi_n}(\mathbf{u})-\mathbf{Q}_\pms(\mathbf{u})\| \to 0 \quad \mbox{a.s.~as $n\to\infty$}. $$  In particular, with probability one, $\widehat{\mathbf{Q}}_{n,\xi_n}(\mathbf{u})\to \mathbf{Q}_\pms(\mathbf{u})$ for every 
$\mathbf{u}\in \mathbb{B}_d\setminus\{\mathbf{0} \}$. 
\end{enumerate}
\end{theorem}
}

\medskip
\noindent\textbf{Proof.} { Part~\textit{(i)} of the proposition  is a trivial consequence of Proposition \ref{propTech}. Turning to  Part~\textit{(ii)},  let $\mathrm{P}\n$ denote 
the empirical distribution  of the sample $\mathbf{X}\n$\!. Then $\mathrm{P}\n\to_w \mathrm{P}_{\X}$ on a probability
one set. Let $\Psi$ be the unique optimal transportation potential from~${\mathrm U}_d$ to~${\mathrm P}_{\X}$ satisfying $\Psi({\bf 0})=0$. 
The assumption that $\mathrm{P}_{\X}\in \mathcal{P}_d$ implies that $\Psi$ is convex and
differentiable at every point of the open unit ball (except, possibly, at the origin) and such that $\nabla \Psi =\mathbf{Q}_\pms$ (we refer to del Barrio et al.~(2018) for details). {Let us assume first that~$\mathrm{P}_{\X}$ has  finite second moments. Then, on a probability
one set, $\mathrm{P}\n$ converges to  $\mathrm{P}_{\X}$ 
and~${\mathrm U}_{d}\n$ converges to  ${\mathrm U}_{d}$ in $L_2$-Wasserstein distance,\footnote{Recall that convergence in $L_2$-Wasserstein distance is equivalent to weak convergence plus convergence of second-order  moments, see Villani~(2009).} 
and it follows from Theorem~2.8 in del Barrio and Loubes~(2019)
that (over the same probability one set) $\Psi_n(\mathbf{u})\to \Psi(\mathbf{u})$ for every~$\mathbf{u}$ in the open unit ball (observe that there is no need to consider
centering constants as in the cited reference, since we have set here $\Psi_n(\mathbf{0})=\Psi(\mathbf{0})=0$). 
Observe that \textit{(i)} implies that also, still with probability one,  $\Psi_{n,\xi_n}(\mathbf{u})\to \Psi(\mathbf{u})$ for every~$\mathbf{u}\in{\mathbb{B}_d}$. Hence, Theorem~25.7 in Rockafellar~(1970) applies, implying  that  
\begin{equation}\label{Psiconv}
\nabla \Psi_{n,\xi_n}(\mathbf{u})=\widehat{\mathbf{Q}}_{n,\xi_n}(\mathbf{u})\to 
\mathbf{Q}_\pms(\mathbf{u})=\nabla \Psi (\mathbf{u})
\end{equation}
 for every $\mathbf{u}\in \mathbb{B}_d\setminus\{\mathbf{0} \}$,   uniformly  over 
compact subsets of the punctured unit ball. This proves \textit{(ii)} under the additional assumption of finite second moment.}

In the absence of a finite second moment for $\mathrm{P}_{\X}$, we still have, with probability one,  that~$\mathrm{P}\n\to_w \mathrm{P}_{\X}$ and 
${\mathrm U}_{d}\n\to_w{\mathrm U}_{d}$  (the last convergence guaranteed by Assumption~\eqref{ConsistencyCondition}). 
Consider the  probabilities (over $\mathbb{R}^d\times\mathbb{R}^d$) $\pi_n := (\nabla \Psi_n \times {\rm Id})\sharp {\mathrm U}_{d}\n$ and $\pi := (\nabla \Psi \times~{\rm Id})\sharp {\mathrm U}_{d}$  induced from~${\mathrm U}_{d}\n$ and~${\mathrm U}_{d}$  through the maps $\nabla \Psi_n \times {\rm Id}$ and~$\nabla \Psi \times {\rm Id}$, respectively. The interpolation property of $\Psi_n$ guarantees that $\pi_n$ has first marginal 
$\mathrm{P}\n$;  similarly, the first marginal of $\pi$ is $\mathrm{P}_{\X}$. Having weakly convergent marginals, the sequence $\pi_n$ is tight. By Lemma 9 in McCann (1995), every convergent subsequence of $\pi_n$ must converge to a joint probability with cyclically monotone support and marginals $\mathrm{P}_{\X}$ and 
${\mathrm U}_{d}$. The only joint probability satisfying these two conditions is $\pi$ and this shows that $\pi_n\to_w \pi$ (on a probability one set). 
This convergence was the only requirement, in the proof of Theorem~2.8 in del Barrio and Loubes~(2019), to conclude that $\Psi_n(\mathbf{u})\to \Psi(\mathbf{u})$ for every~$\mathbf{u}\in{\mathbb{B}_d}$. The same Theorem~25.7 of  Rockafellar~(1970) thus applies, yielding~\eqref{Psiconv}. This completes the proof. \qed

Note that  the first statement in Theorem 2.2  does not entail {\it uniform} convergence of~$\Psi_{n,\xi_n}$ to $\Psi$ over $\mathbb{B}_d$. Pointwise convergence, however, follows from the proof of part~{\it (ii)}.

\subsection{Numerical illustration}

In this section, we provide some numerical illustrations of the unsmoothed (${\bf Q}_n$) and smoothed ($\widehat{\bf Q}_{n,\xi}$) empirical quantile functions for i.i.d.~samples of size $n=1000$ from  the bivariate ${\cal N}({\bf 0}, {\bf I})$, spherical~$t_3$, and elliptical hyperbolic\footnote{Recall that a random vector $\X$ admits an elliptical hyperbolic distributions if $\X=_d \left(  \gamma D \right)^{-1/2}{\bf Y}$ where~$D$ is $\chi_{1/\gamma}^2$-distributed,  independent of ${\bf Y} \sim \mathcal{N}_d ( {\bf 0}, {\boldsymbol\Sigma})$, and ${\boldsymbol\Sigma}$ a symmetric~$d\times d$ correlation matrix with   off-diagonal elements  0.5.}  ($\gamma~=~1/3$) distributions. 
The same seeds were used in all these simulations in order to illustrate how  increasingly outlying observations influence the graphs under increasingly heavier tail densities.

The grids $\{{\bf w}_1, \ldots, {\bf w}_n\}$  (of the random $\mathfrak{w}\n$ type) underlying    bivariate numerical illustrations were obtained 
via a polar coordinate construction 
$${\bf w}_i :=  {i \over n+1}
 \big(
\cos (\phi_i) ,
\sin (\phi_i)  \big)^\prime
,\qquad i=1,\ldots,n$$
 with~$\phi_{1},\ldots , \phi_{n}$ independently and uniformly distributed over~$[0,2\pi)$.  
For $d \geq 3$ (Sections~4 and 5),  we simulated $d$-dimensional i.i.d.~${\cal N}({\bf 0},{\bf I})$ $n$-tuples~${\bf Z}_1,\ldots, {\bf Z}_n$ and let  
\begin{equation}
{\bf w}_i :=  {{\bf Z}_i \over \| {\bf Z}_i \|} \widehat{F}\n_{R} (\| {\bf Z}_i \|), \quad  i= 1,\ldots,n,
\label{Qell_emp}
\end{equation} 
where   $\widehat{F}_{R}\n $ stands for  the empirical distribution function of  the moduli $\| {\bf Z}_i \|$. In all simulations and illustrations we used $\log \xi=300$.

\begin{figure}[htbp]
\centering
\includegraphics[width=50mm,trim=.5cm 1cm .5cm 0cm,clip]{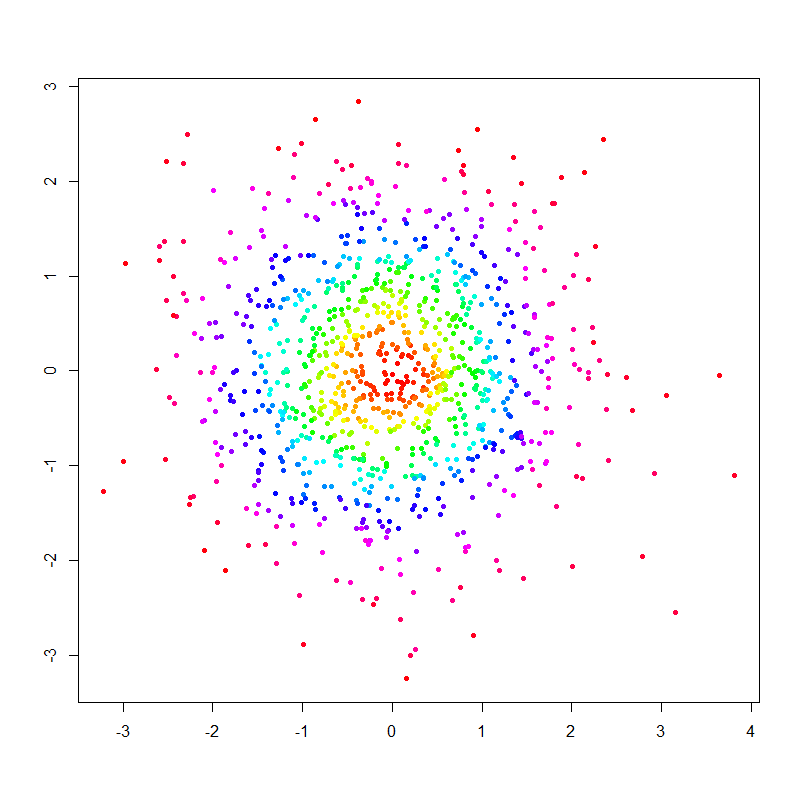}
\includegraphics[width=50mm,trim=.5cm 1cm .5cm 0cm,clip]{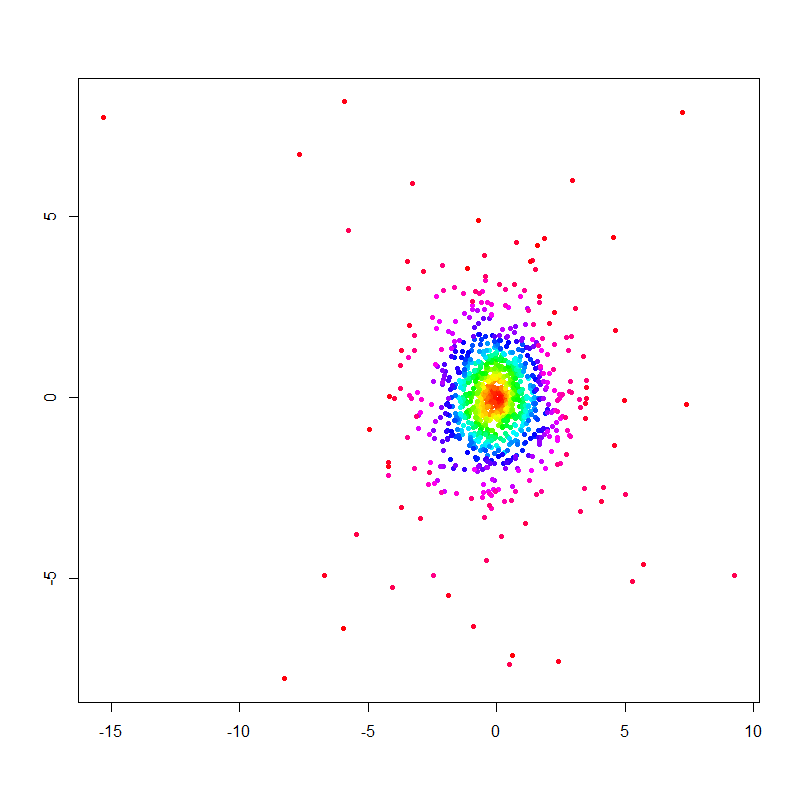}
\includegraphics[width=50mm,trim=.5cm 1cm .5cm 0cm,clip]{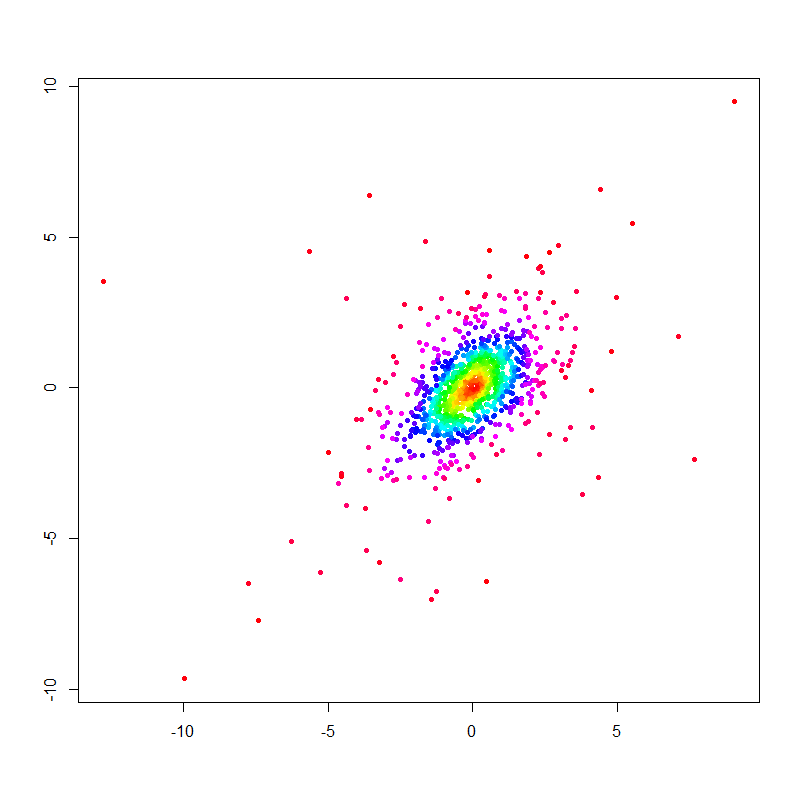}
\includegraphics[width=50mm,trim=.5cm 1cm .5cm 0cm,clip]{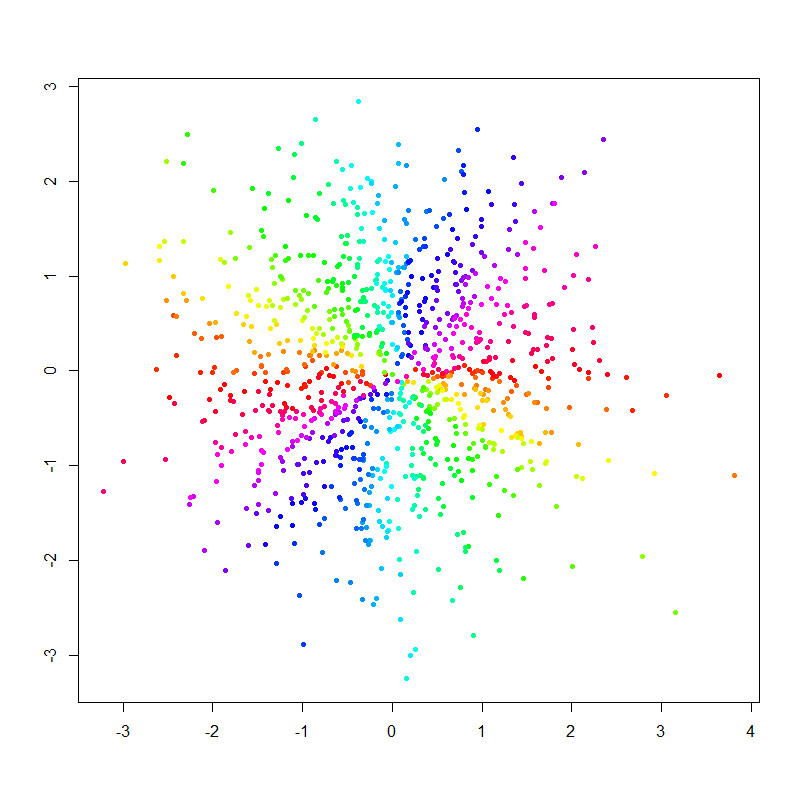}
\includegraphics[width=50mm,trim=.5cm 1cm .5cm 0cm,clip]{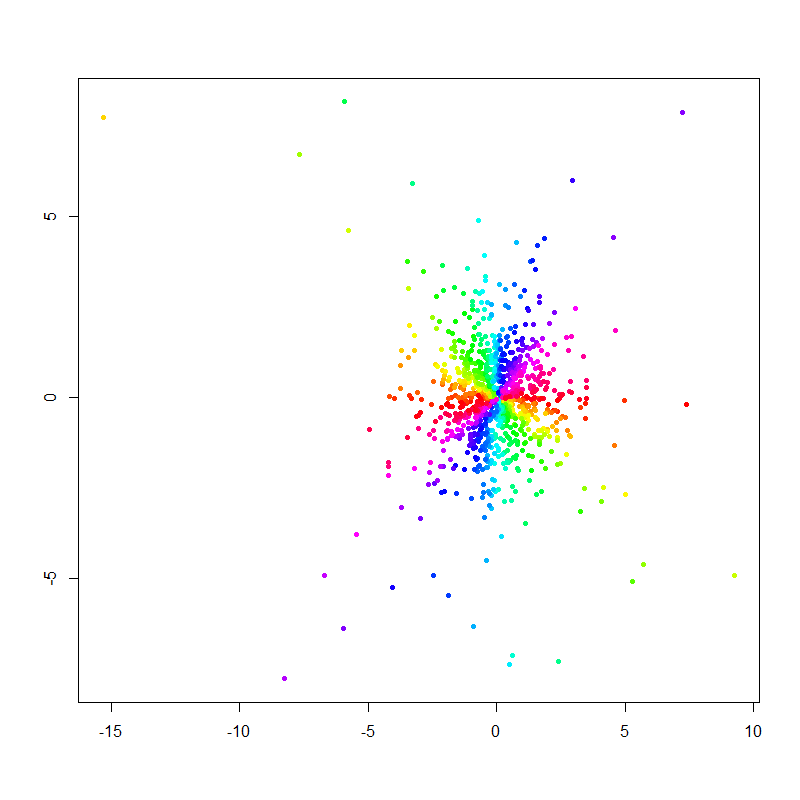}
\includegraphics[width=50mm,trim=.5cm 1cm .5cm 0cm,clip]{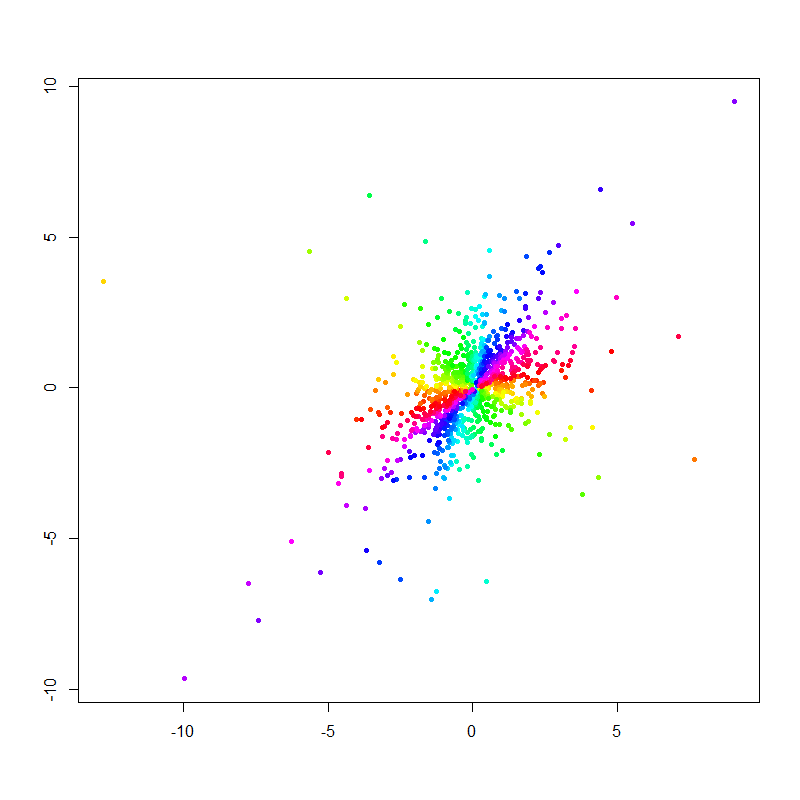}
\caption{The optimal couplings ${\bf Q}_n = {\bf Q}_\pm\n$ between samples from bi\-va\-riate~${\cal N}({\bf 0}, {\bf I})$ (left), spherical $t_{3}$ (center) and elliptical hyperbolic  (with para\-meter~$\gamma~=~1/3$)  (right) distributions   and a random grid ${\mathfrak{w}}\n$;   colours are used to visualize the empirical quantile contours (top) and empirical signs (bottom). Each sample is of size $n=1000$; one replication,~$m~\!=~\!1$ random grid.}
\label{coupling}
\end{figure}


 \begin{figure}[htbp]
\centering
\includegraphics[width=50mm,trim=.5cm 1cm .5cm 0cm,clip]{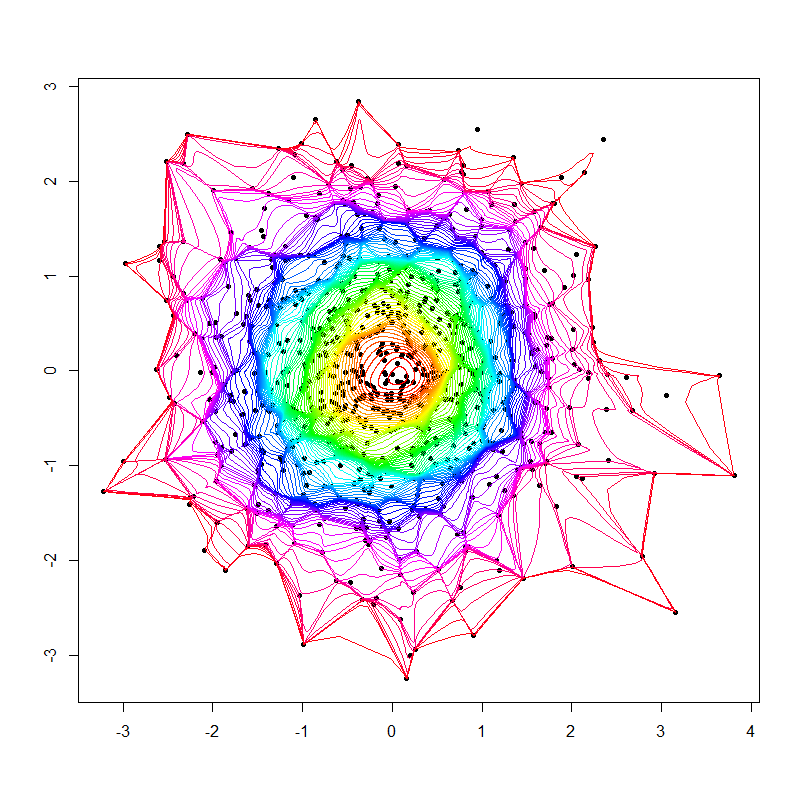}
\includegraphics[width=50mm,trim=.5cm 1cm .5cm 0cm,clip]{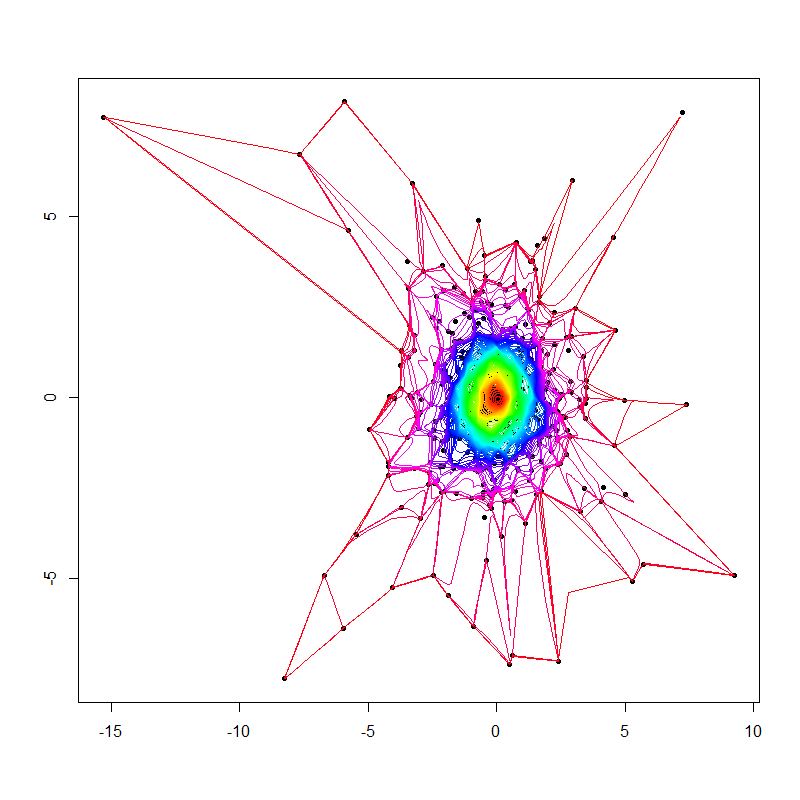}
\includegraphics[width=50mm,trim=.5cm 1cm .5cm 0cm,clip]{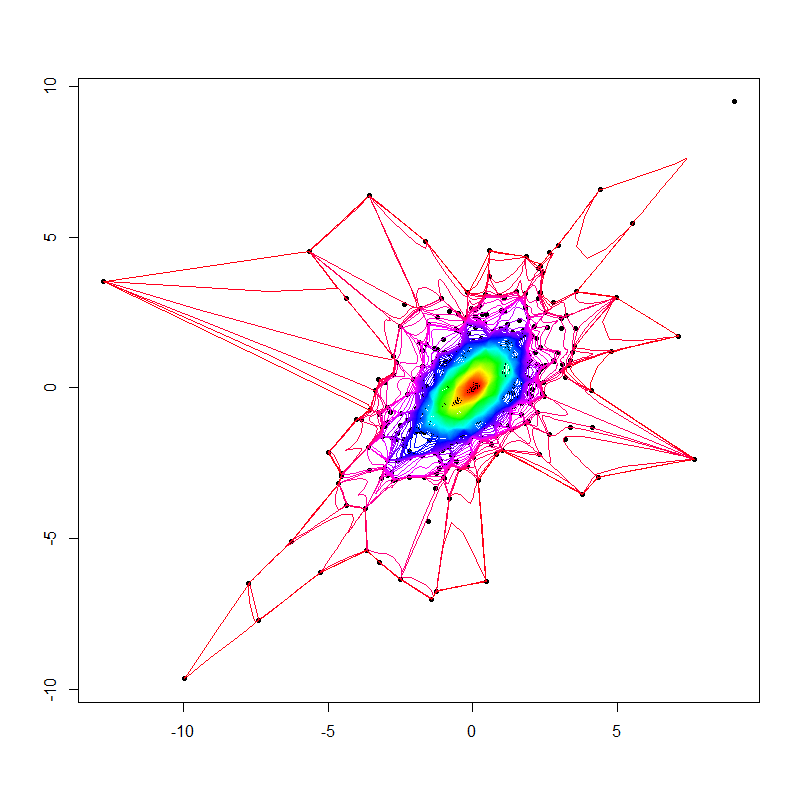}
\includegraphics[width=50mm,trim=.5cm 1cm .5cm 0cm,clip]{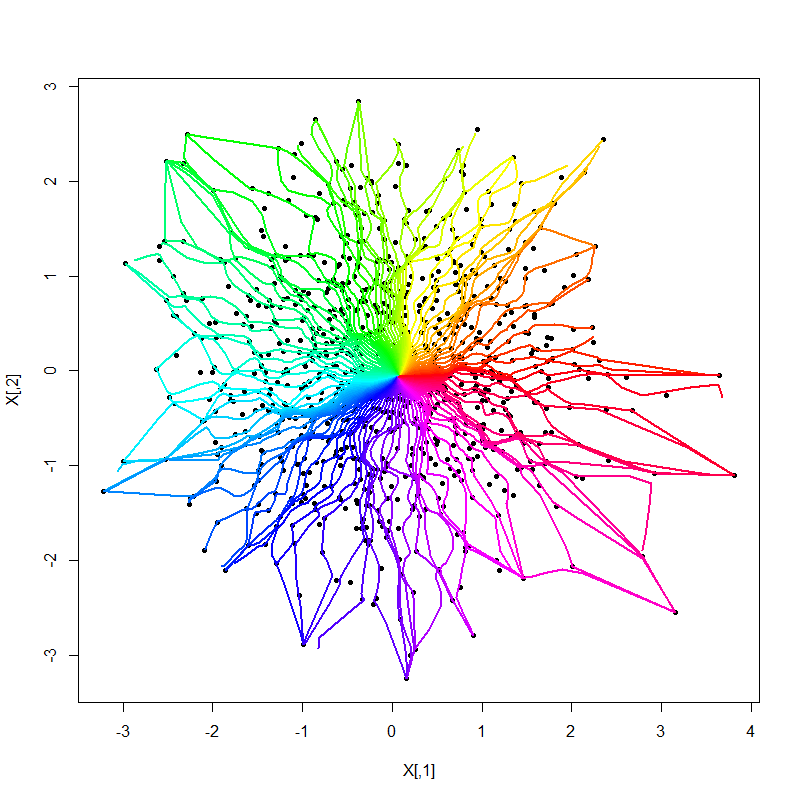}
\includegraphics[width=50mm,trim=.5cm 1cm .5cm 0cm,clip]{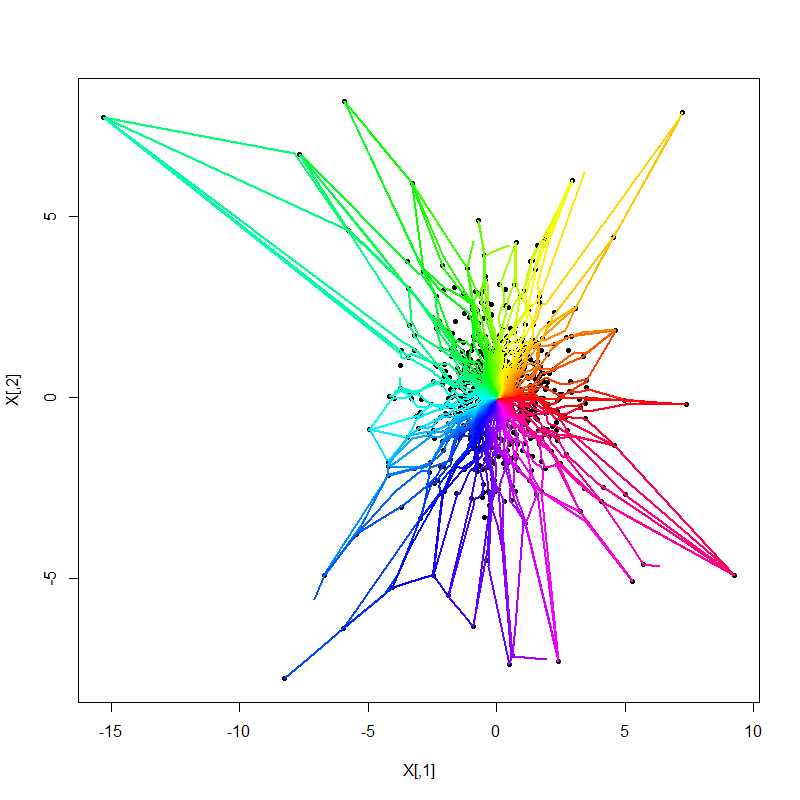}
\includegraphics[width=50mm,trim=.5cm 1cm .5cm 0cm,clip]{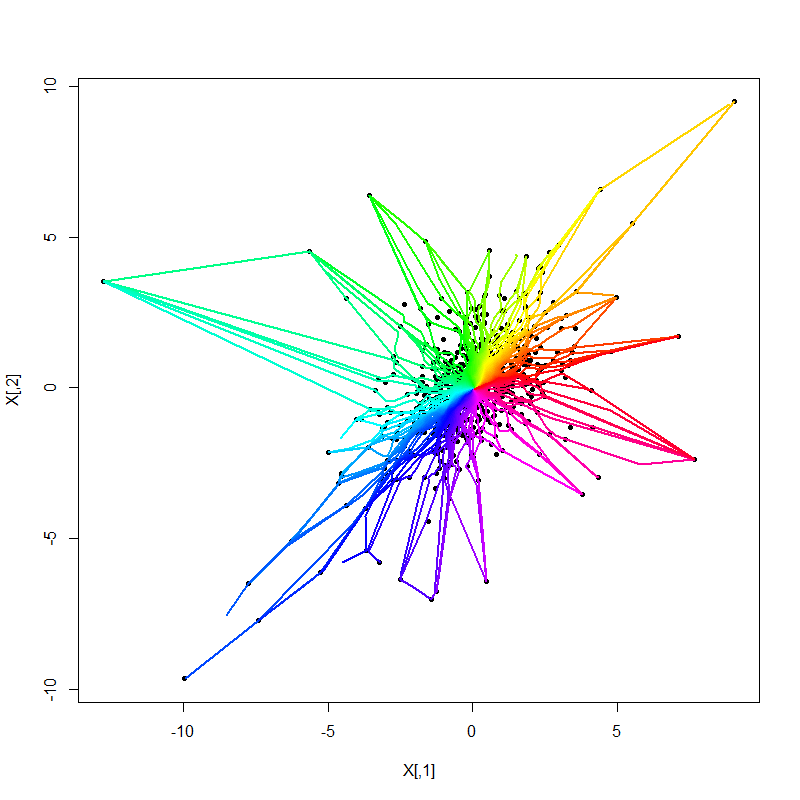}
\caption{The smoothed empirical quantile contours (top) and sign curves (bottom) based on $\widehat{\bf Q}_{n,\xi}$, $\xi=\exp(300)$, for the same samples as in Figure~1, averaged over $m=10$ random grids. }
\label{coupling'}
\end{figure}

\begin{figure}[htbp]
\centering
\includegraphics[width=130mm,trim=1cm 2cm 1cm 0cm,clip]{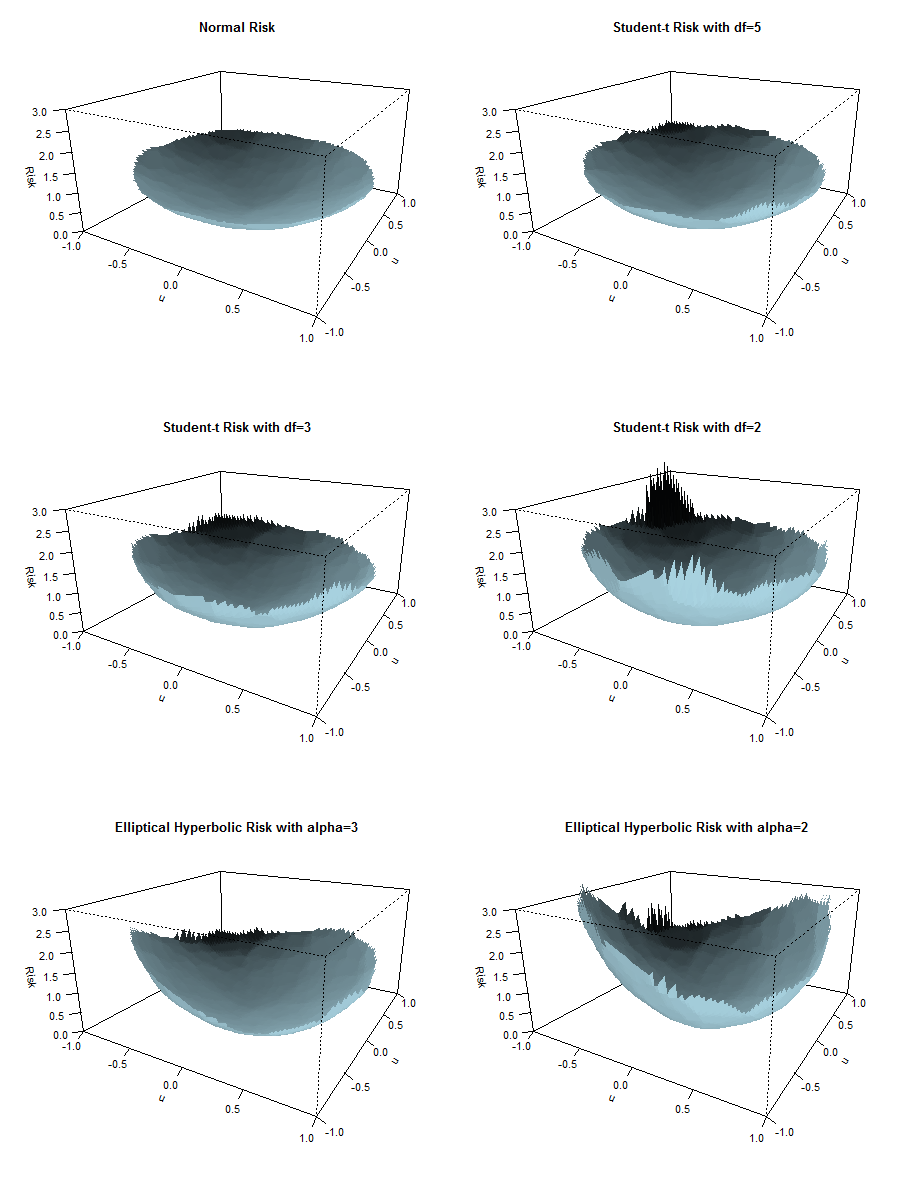}
\caption{Empirical potential surfaces $ \Psi _{n,\xi} ({\bf u})$ ($\xi=\exp(300)$)  for the same samples as in as  in Figures 1 and 2 (normal, spherical $t_3$, elliptical hyperbolic with~$1/\gamma =3$), spherical $t_5$ and $t_2$, and elliptical hyperbolic with~$1/\gamma= 2$. Each sample is of size $1000$; $m=10$ random grids.}
\label{potential_t_normal}
\end{figure}

 In Figure 1, the unsmoothed empirical quantiles ${\bf Q}_n $, providing the optimal couplings between the three samples and one single random grid $\mathfrak{w}\n$ (i.e.~$m=1$)   are shown.  In the top panels, the $n$ observations $\X_i$ are given   colors    ranging, center outward, from  red, yellow, green, blue, to cyan according to their empirical quantile orders~$\Vert {\bf F}_n(\X_i)\Vert$.   The same color code is used clockwise in the bottom panels   to visualize the~$n$ signs~$ {\bf F}_n (\X_i)/\Vert {\bf F}_n (\X_i)\Vert$.
 
Figure~2 provides, for the same samples, the smoothed contours\footnote{The  quantile contour of order $p$ is defined as $C_{n,\xi,p} :=\{  \widehat{\bf Q}_{n,\xi}({\bf u}) \,| \, \| {\bf u}\| = p \}$, $p \in (0,1)$} and sign curves\footnote{A sign curve is defined as $\{\widehat{\bf Q}_{n,\xi}(c{\bf s})\vert c\in [0,1) \},  \  {\bf s}\in{\mathbb{S}}_{d-1}$; while quantile contours are strictly nested, sign curves {\it never} intersect.} corresponding to $\widehat{\bf Q}_{n,\xi}$ with $m=10$.  Note that, for the heavy-tailed $t$ and   hyperbolic distributions, the outer contours, containing  the most outlying observations, exhibit  irregular star shapes.  As for the sign curves running through extreme observations, although they never intersect,  they seem to join when reaching  those outlying observations. This is a consequence of the irregular distribution of the corresponding directions which, for two ``angularly consecutive outliers,"  and after averaging over $m$ grids, can be either quite big or quite small---so small that  limited-resolution pictures fails to separate them, creating bundles. 

In Figure 3, we plot the smoothed empirical potential surfaces ${\bf u} \mapsto \Psi _{n,\xi} ({\bf u})$ for the same samples (normal, elliptical $t_3$, elliptical hyperbolic $\gamma =1/3$) as  in Figures 1 and~2, next to spherical $t_5$ and $t_2$, and elliptical hyperbolic with $\gamma= 1/2$. Note that with increasing tail weight (i.e., decreasing degrees of freedom or increasing $\gamma$ values), the surfaces get steeper   near the border of the unit ball $\mathbb{B}_2$.

\section{Risk measurement based on $\Psi_{n,\xi}$ and $\widehat{\bf Q}_{n,\xi}$}

Let $X$ denote a real-valued risk, with traditional distribution and quantile  functions~$F$ and~$Q$, respectively.  Gushchin and Borzykh (2018)  emphasize the fundamental role,     in a variety of univariate   risk measurement problems, of  the integrated distribution and quantile functions 
$$\Phi(x):= \int_{-\infty}^xF(z)\, dz\quad\text{ and }\quad  \Psi(p):=\int _0^pQ(v)\, dv, $$ provided that the second integral exists.  Concentrating further on integrated quantiles, 
we refer to Kusuoka~(2001) who introduced the  quantity (the reason for the notation, which is ours, will appear later on)
$$\varrho^X_{{\mathrm U}(0,1)} := \int_0^1 v \, Q(v) \, dv = {\mathrm E}[V\, Q(V)] = \text{Cov}(V, Q(V)) + \frac{1}{2}{\mathrm E}X,$$ where  $V$ is uniform over $[0,1)$,  as a  regular coherent risk measure (provided that the integral---equivalently,  ${\mathrm E}X$---exists). This $\varrho^X_{{\mathrm U}(0,1)}$ is then estimated by 
$$\hat{\varrho}^X_{n,{\mathrm U}(0,1)}:=\frac{1}{n}\sum_{i=1}^n\frac{i}{n}X_{(i)}= \frac{1}{n}\sum_{i=1}^n\frac{R\n_i}{n}X_{i}$$
 where $X_{(i)}$ and $R\n_i$ stand for the $i$th order statistic and the rank of $X_i$, respectively,  in an i.i.d.~sample $X_1,\ldots,X_n$ of size $n$. 

 Since, for $d=1$,~$F_\pms (x) =2F(x)-1$, we have $Q_\pms (u)= Q((1+u)/2)$, elementary calculation yields 
$$ 4\varrho^X_{{\mathrm U}(0,1)} = 4\int_0^1\! v Q(v) \, dv = \int_{-1}^1\! uQ_\pms (u)\, du +   \int_{-1}^1 \! Q_\pms  (u)\, du 
=\int_{-1}^1\! uQ_\pms (u)\, du +  2{\mathrm E}X
$$
reducing,  for centered $X$, to 
 $
\int_{-1}^1\! uQ_\pms (u)\, du
$. 
We thus quite naturally   consider 
$$\varrho^{\bf X}_{{\mathrm U}_d} := \int_{\mathbb{B}_d}{\bf u}^\prime {\bf Q}_\pms ({\bf u})\, d {\mathrm U}_d({\bf u})$$ as a measure of the risk of ${\bf X}-{\mathrm E}{\bf X}$. 


This risk measure 
 is closely related to another class of measures of risk, the {\it maximal correlation risk measures}\footnote{The terminology ``correlation" in this context is somewhat improper as  $\rho_\mu$ is a  covariance, that can take values higher than one. Moreover, it measures the risk of ${\bf X}-{\mathrm E}{\bf X}$ rather than the risk of $\bf X$.}  developed by R\"uschendorf (2006) and Ekeland et al.~(2012). 
The  {\it maximal correlation risk measure with respect to the baseline distri\-bution~$\mu$} of a $d$-dimensional risk $\bf X$ is defined as 
$$
\rho_\mu^{\X} :=\sup \{ \mathrm{E}( \langle \X , \tilde{\bf U}\rangle ): \tilde{\bf U} \sim \mu \}.
$$
By choosing the baseline distribution as the uniform distribution over the unit ball ${\mathbb{B}_{d}}$, i.e., letting  $\mu={\mathrm U}_d $,    it follows from Appendix~B in Ekeland et al.~(2012) that the maximal correlation risk measure  is 
\begin{align}
\rho^{\X}:= \rho_{{\mathrm U}_d}^{\X} & = \mathrm{E} \left( \langle {\bf U} , {\bf Q}_\pms ({\bf U}) \rangle \right) = \mathrm{E} \left( \langle {\bf X} , {\bf F}_\pms ({\bf X}) \rangle \right)  
=\int_{\mathbb{B}_d}{\bf u}^\prime {\bf Q}_\pms ({\bf u})\, d {\mathrm U}_d({\bf u}) = 4\varrho^{\bf X}_{{\mathrm U}_d}
\nonumber 
\end{align}
(whenever the expectation exists): $\rho^{\X}:= \rho_{{\mathrm U}_d}^{\X}$ and $\varrho^{\bf X}_{{\mathrm U}_d}$ thus coincide up to a factor 4.

The  maximal correlation  $\rho^{\X}$ can  be estimated   by 
\begin{align}
 \widehat \rho_{n}^{\X} := \frac{1}{n} \sum_{i=1}^n   \, \langle {\bf u}_{i}, \widehat{\bf Q}_{n,\xi} ({\bf u}_{i}) \rangle . 
\label{risk}
\end{align}
The graph of~${\bf u}\mapsto \langle {\bf u}, {\bf Q}_\pms ({\bf u})\rangle$ is a surface in $\mathbb{R}^{d+1}$---call it  the {\it theoretical risk surface}---consistently estimated by the {\it empirical risk sur\-face} 
\begin{align}
{\bf u}  \mapsto \langle {\bf u}, \widehat{\bf Q}_{n, \xi} ({\bf u}) \rangle , \hspace{3mm}  {\bf u} \in \mathbb{B}_{d} \setminus \{{\bf 0}\} \label{potential}
\end{align}
and $\rho^{\X}$ ($\widehat \rho_{n}^{\X}$) is the average height of that (empirical) risk surface. 
Note that from Theorem~2.2$(ii)$ it follows that for every ${\bf u} \in \mathbb{B}_{d} \setminus \{{\bf 0}\}$
$$
 \langle {\bf u}, \widehat{\bf Q}_{n, \xi} ({\bf u}) \rangle \to \langle {\bf u}, {\bf Q}_{\pms} ({\bf u}) \rangle \mbox{ a.s. as  } n\to \infty. 
$$
The empirical risk surfaces for the samples considered in Figures 1--3 are shown in Figure~4.\\

\begin{figure}[h!]
\centering
\includegraphics[width=130mm,trim=1cm 2cm 1cm 0cm,clip]{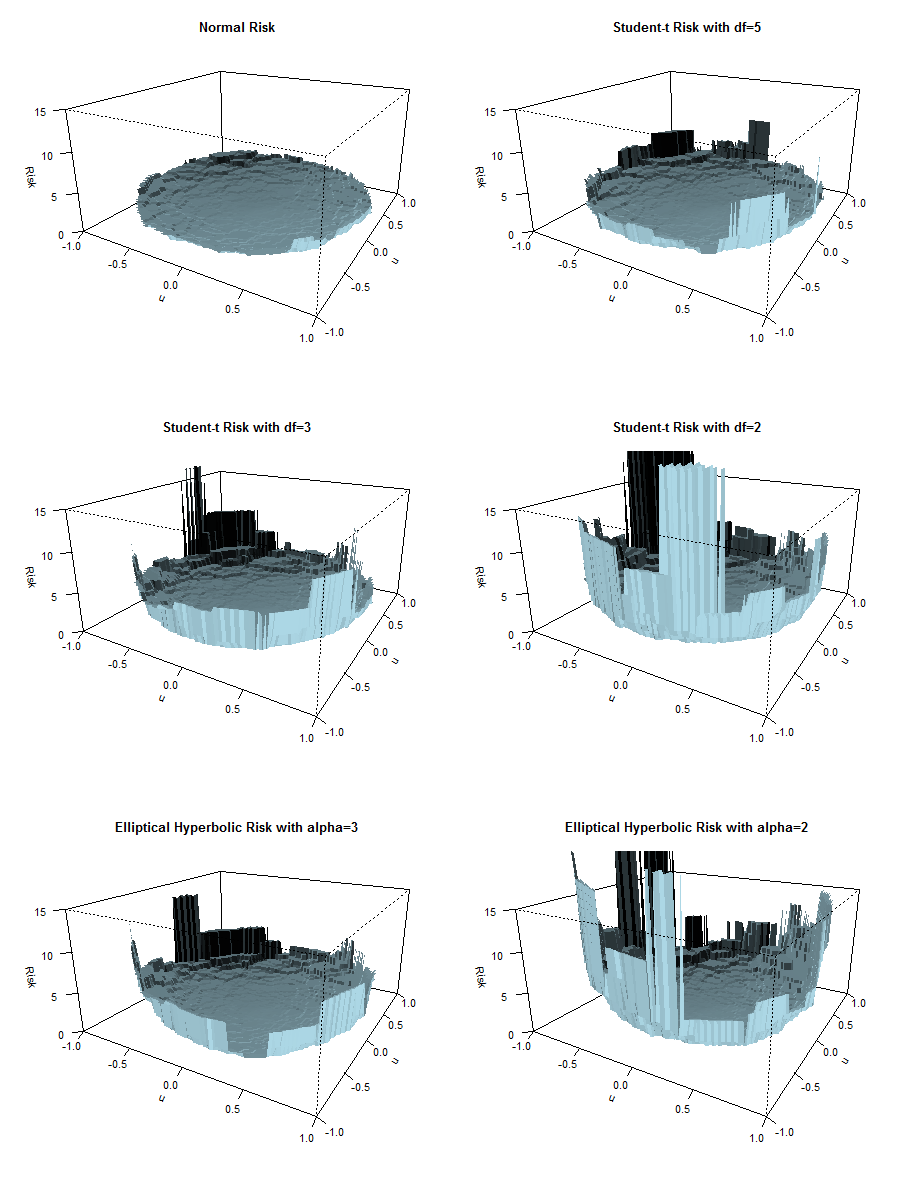}
\caption{Risk surfaces of (centered) spherical bivariate normal (top left),~$t_{5}$ (top right), $t_{3}$ (center left), $t_{2}$ (center right), and   bivariate elliptical hyperbolic   with $1/\gamma=3$ (bottom left) and $1/\gamma=2$ (bottom right)   samples. Each sample has size $n=1000$.}
\label{risk_t_normal}
\end{figure}

\begin{table}[ht]
\caption{Summary of $\hat{\rho}_{n}$ values averaged over 100 samples of   dimensions~$d=2,\ldots,5$ and  size $n=1000$ from spherical bivariate normal,  $t_{5}$, and $t_{3}$ distributions.}
\centering
\begin{tabular}{c c c c }
\hline \hline \vspace{-2mm}\\\vspace{1mm}
 $d$	& Normal 		& $t_5$	& $t_3$ 	 \\
\hline
2		& 0.80		& 1.00	& 1.21	\\
3		& 0.97		& 1.23	& 1.48 	 \\
4		& 1.08		& 1.37	& 1.66	 \\
5		& 1.16		& 1.47	& 1.77 \vspace{1mm}	\\ \hline \hline
\end{tabular}
\end{table}

  In Table 1, the values of $\widehat{\rho}_{n}^{\X}$ are given for simulated samples of standard multivariate normal and $t$ distributions for different dimensions. In each case, an average  is taken over~100~samples of size $n=1000$. Note how  $\widehat{\rho}_{n}^{\X}$ (hence the estimated risk) increases with  dimension for fixed tailweight and with tailweight for fixed dimension.

\vspace{0.3cm}The contribution to $\rho^{{\bf X}}$ of  tail events and outlying observations can be amplified or reduced by restricting the baseline uniform ${\mathrm U}_d$ 
 to   some chosen quantile region, that is, by considering, for $p\in (0,1)$, the {\it maximal tail correlation risk measure} 
\begin{align*}
\rho_p^{{\X},+} 	& :=  \mathrm{E}\left( \langle {\bf U}, {\bf Q}_{\pms} ({\bf U}) \rangle \, | \, \|{\bf U}\|> 1-p\right) =\frac{1}{p}\int_{{\mathbb B}_d\setminus (1-p){\mathbb B}_d}{\bf u}^\prime {\bf Q}_\pms ({\bf u})\, d {\mathrm U}_d({\bf u})
\end{align*}
which can be estimated  by 
\begin{align}
 \widehat \rho^{{\X},+}_{n,p} & :=  \frac{1}{\sum_{i=1}^n 1_{[\|{\bf u }_i \|> 1-p]}}  \sum_{i=1}^n   \, \langle {\bf u}_{i},\widehat{\bf Q}_{n,\xi} ({\bf u}_{i}) \rangle \, 1_{[\|{\bf u}_{i}\|> 1-p]}, \nonumber
\end{align}
or the {\it maximal trimmed correlation risk measure} 
\begin{align*}
\rho_p^{{\X},-}  	& :=  \mathrm{E}\left( \langle {\bf U}, {\bf Q}_{\pms} ({\bf U}) \rangle \, | \, \|{\bf U}\| \leq 1-p\right)  =\frac{1}{1-p}\int_{ (1-p){\mathbb B}_d}{\bf u}^\prime {\bf Q}_\pms ({\bf u})\, d {\mathrm U}_d({\bf u})
\end{align*}
 estimated  by 
\begin{align}
 \widehat \rho^{{\X},-}_{n,p}  & :=  \frac{1}{\sum_{i=1}^n 1_{[\|{\bf u }_i \|\leq 1-p]}}  \sum_{i=1}^n   \, \langle {\bf u}_{i},\widehat{\bf Q}_{n,\xi} ({\bf u}_{i}) \rangle \, 1_{[\|{\bf u}_{i}\|\leq 1-p]}. \nonumber
\end{align}
Clearly, $\rho^{{\X},+}_{p}$ and $\rho^{{\X},-}_{p}$ ($ \widehat \rho^{{\X},+}_{n,p}$ and $ \widehat \rho^{{\X},-}_{n,p}$) also provide an evaluation of the respective contributions of tail and central regions to the risk, which decomposes into
$$ \rho^{\X} = p\rho^{{\X},+}_{p}  + (1-p) \rho^{{\X},-}_{n,p} \qquad \big(\widehat \rho_{n,p}^{\X}= p\widehat \rho^{{\X},+}_{n,p} + (1-p)\widehat \rho^{{\X},-}_{n,p}
\big).
$$ 

Table~2 shows, for $p=0.05$, the values of $\widehat{\rho}^{{\X},+}_{n,0.05}$ (tail risk) and  $\widehat{\rho}^{{\X},-}_{n,0.05}$ (trimmed risk), along with their relative contributions to the global risk $\rho^{\X}$ for the same distributions as in Table~1. Both risks are increasing with tailweight (fixed dimension) and  the dimension~$d$ (fixed tailweight), just as the global risk in Table~1. For fixed tailweight, the relative contribution of  tail risk is increasing, and  the relative contribution of  trimmed risk is decreasing with $d$. However, for fixed dimension, the relative contribution of  tail risk is decreasing with tail weight (that of  trimmed risk is increasing).


\begin{table}[ht]
\caption{Summary of $\widehat{\rho}^{{\X},+}_{n,0.05}$ (tail risk) and  $\widehat{\rho}^{{\X},-}_{n,0.05}$ (trimmed risk) values  averaged over 100 samples of   dimensions~$d=2,\ldots,5$ and  size $n=1000$;  in parentheses, their relative contributions to $\widehat{\rho}_n^{\X}$. } 
\centering
\begin{tabular}{c c c c c c c c }
\hline \hline \vspace{-2mm}\\ \vspace{1mm}
 $d$&&$\widehat{\rho}^{{\X},+}_{n,0.05}$& & &&$\widehat{\rho}^{{\X},-}_{n,0.05}$&  \\ \hline \\
	& Normal 		& $t_5$	& $t_3$ & \ &Normal 		& $t_5$	& $t_3$ 	\\
\hline
2		&2.61\,\ (84\%)		& 4.17\ (79\%)		& 6.30    \ (74\%) & \ &0.70 \ (16\%)&0.83\ (21\%)	&0.94\ (26\%)	\\
3		&2.69 \ (86\%)	         & 4.37\ (82\%)		& 6.48\ (78\%)		& \ &0.88 \ (14\%)&1.06\ (18\%)	& 1.22\ (22\%)\\
4		& 2.73 \ (87\%)		& 4.30\ (84\%)		& 6.35  \ (81\%)	              & \ &0.99 \ (13\%)&1.22\ (16\%)	&1.41\ (19\%)	 \\
5		& 2.78 \ (88\%)		& 4.35\ (85\%)		& 6.17  \ (83\%)	              & \: &1.07 \ (12\%)&1.32\ (15\%)	&1.54\ (17\%)	 \\ \hline
\end{tabular}
\end{table}
\color{black}


\section{Quantile plots based on the volumes of center-outward quantile regions} 

{\it QQ plots} are  another fundamental tool in the analysis of  risk which, due to the absence of a canonical quantile concept in $\mathbb{R}^d$,  remains essentially limited to the univariate context.  In this section, we introduce the notions of  a {\it center-outward quantile volume} and {\it center-outward   QQ plot}   based on  the   volumes $V(p):=\int \ldots \int_{{\bf Q}_{\pms}(p \mathbb{S}_{d-1})} \prod_{i=1}^d dx_i$, $p\in [0,1)$ of the center-outward quantile regions and their empirical counterparts. 

As a justification of the concept, consider the particulat case of an elliptical risk. A closed form then is easily obtained for the volumes $V(p)$: it follows from \eqref{Fell} that 
\begin{eqnarray}
 V(p) &=& | {\boldsymbol\Sigma} |^{1/2}\int \ldots \int_{\{{\bf z}= {\boldsymbol\Sigma}^{-1/2}{\bf Q}_{\pms} ({\bf u}): \| {\bf u}\| \leq p \}}\prod_{i=1}^d dz_i \nonumber \\
 &=& { | {\boldsymbol\Sigma} |^{1/2}\over d}
 \int_0^p \int_0^{2\pi} \int_0^{\pi} \ldots \int_0^{\pi} \,
 \prod_{j=1}^{d-1} \sin^{d-1-j} \phi_j \,\, dr_z^d \, d\phi_1 \, d\phi_2 \ldots d\phi_{d-1} \nonumber \\
 &=& { | {\boldsymbol\Sigma} |^{1/2}\over d}
 \int_0^p \int_0^{2\pi} \int_0^\pi \ldots \int_0^{\pi} \,
 \prod_{j=1}^{d-1} \sin^{d-1-j} \phi_j \,\, dQ^d_R(r_u) \, d\phi_1 \, d\phi_2 \ldots d\phi_{d-1} \nonumber \\
 &=& | {\boldsymbol\Sigma} |^{1/2} \, {\pi^{d/2} \over \Gamma (1+d/2)}
  Q^d_R(p), \label{elliptical}
\end{eqnarray}
where $Q_R$ denotes the quantile function of the distribution of $\|{\bf Z} \|$. Note that in the multivariate Gaussian case $\|{\bf Z} \|^2$  is $\chi^2_d$ distributed so that $Q^2_R$ equals the quantile function of the $\chi^2_d$ distribution.

Explicit expressions of $ V(p) $ for general distributions seem more tricky---analytic forms of non-elliptical center-outward quantile functions moreover are hardly available, as they would require solving the corresponding Monge-Amp\`ere equations. Estimating  $V(p)$, however, is possible, and relatively easy via the volumes $V_{n,\xi} (p)$ of  the empirical  quantile regions   
$$\bar{R}_{n,\xi,p} :=\{  \widehat{\bf Q}_{n,\xi}({\bf u}) \,| \, \| {\bf u}\|\leq p \}, \qquad p \in (0,1).$$  
  Using polar coordinates, indeed, we obtain
\begin{align*}
V_{n,\xi}(p)	& = \int _{\bar{R}_{n,\xi,p}} dx_1 \, dx_2 \ldots dx_d 
 	 = \int _{{\bf u} \in p\mathbb{B}_d }  J_{n,\xi}({\bf u}) \, du_1 \, du_2 \ldots du_d \\
 	& = \int_0^p \int_0^{2\pi} \int_0^\pi \ldots \int_0^{\pi}  \,  J_{n,\xi}( r \, {\bf v}( {\boldsymbol \phi}) )  \, \left\{ r^{d-1} \, \prod_{j=2}^{d-1} \sin^{d-j} \phi_j  \right\} \, d\phi_1 \, d\phi_2 \ldots d\phi_{d-1} \, dr 
\end{align*}
with  
\begin{align*}
{\boldsymbol \phi}:= \left[
\begin{array}{c} \phi_1\\ \vdots\\ \phi_{d-1}\end{array}
 \right] 
&,\qquad 
{\bf v}( {\boldsymbol \phi}) := \left[ \begin{array}{l} \cos \phi_{1} \\ \sin \phi_{1} \cos \phi_{2}  \\ \sin \phi_{1} \sin \phi_{2} \cos \phi_{3} \\ \vdots \\ \sin \phi_{1} \dots \sin\phi_{d-2} \cos \phi_{d-1} \\  \sin \phi_{1} \dots \sin\phi_{d-2} \sin \phi_{d-1} \end{array}\right], 
\end{align*}
and the Jacobian  
\begin{align*}
J_{n,\xi}({\bf u}) &:=  \left| \nabla \widehat{\bf Q}_{n,\xi}({\bf u}) \right| 
  = \xi^d \,  \left|  \sum_{i=1}^n  w_{i, \xi}\left( {\bf u} \right)  \left(\X_i-\overline{\X} \right) \X_i^\prime \right|  \\ 
 &  = \xi^d \, \left|\left(\mathbf{X}^{(n)}-\overline{\X}{\bf 1}_n^\prime \right) {\bf W}_\xi({\bf u})\left(\mathbf{X}^{(n)}-\overline{\X}{\bf 1}_n^\prime \right)^\prime  \right| ,
 \end{align*}
where ${\bf W}_\xi({\bf u}) :=  \text{diag}\left\{  w_{1, \xi}\left( {\bf u} \right),   \dots,  w_{n, \xi}\left( {\bf u} \right)  \right\}$, with weights $w_{i, \xi}\left( {\bf u} \right)$ taken from \eqref{sven},  
$$\mathbf{X}\n=\left( \X_1, \ldots  \X_n\right) =:(X_{ij})^\prime_{1\leq i\leq n, 1\leq j \leq d}, \quad 
\bar{X}_j := \sum_{i=1}^n \, w_{i, \xi}\left( {\bf u} \right)  \, X_{ij},$$
 and 
$\overline{\X} := \left( \bar{X}_1, \ldots  \bar{X}_d\right)^\prime$. Note that  $\left(\mathbf{X}^{(n)}-\overline{\X}{\bf 1}_d^\prime \right) {\bf W}_\xi({\bf u})\left(\mathbf{X}^{(n)}-\overline{\X}{\bf 1}_d^\prime \right)^\prime $ is a weighted covariance matrix.

A heuristic goodness-of-fit test can be performed by comparing, via eye-inspection of  (a discretized version of) the plot of $ V_{n,\xi}(p)$ against the volumes $V(p)$   associated with some reference distribution.   In case of an i.i.d.\  multivariate normal sample, for instance,  the  empirical volumes $V_{n,\xi} (p)$  based on \eqref{elliptical} 
are expected to be  linearly  related with the corresponding quantiles  of the $(\chi^2_d)^{d/2}$ distribution. Hence, a plot of the empirical volume quantiles $V_{n,\xi} \big(i /( n+1)\big)$ ($i=1,\ldots,n$) against the latter should  not  ``significantly deviate'' from the main diagonal of the unit square.  This provides an informal graphical goodness-of-fit test for multivariate normality   in the same spirit as  the generalized QQ plots introduced in Beirlant et al.~(1999) based on the  generalized quantiles\footnote{Those  generalized quantiles refer to the volumes of the smallest ellipsoids (or other type of contours) which contain at least $100p \%$ of the data.} 
  proposed by Einmahl and Mason~(1992). 
  
  Similarly,  a linear relationship should be visible, in case of 
heavy-tailed distributions, near the top volumes ($p\approx 1$) in the graph of $-\log (1-{i /( n+1)})\mapsto\log V_{n,\xi} (1-{i /( n+1)})$   ($i=1,\ldots,n$), see also Beirlant et al.~(1999).
  However, for heavy tailed distributions the star  shape of the top contours as detected in Figure 2  entails a serious downward deflection near the highest volumes, which hampers its use for statistical purposes.     
However,  the generalized median $V(0.5)$ or other central volumes    can be used as robust risk indicators. 

In Figure \ref{multivariate_normal_gof}, 
the $(20j)$-th smallest volumes with $j=1,\ldots,49$, 
from a 2 and 3-dimensional standard normal sample of size $n=1000$ are plotted against the theoretical values $V({20j \over 1001})$ following \eqref{elliptical}.

\begin{figure}[hh]
\centering
\includegraphics[width=120mm]{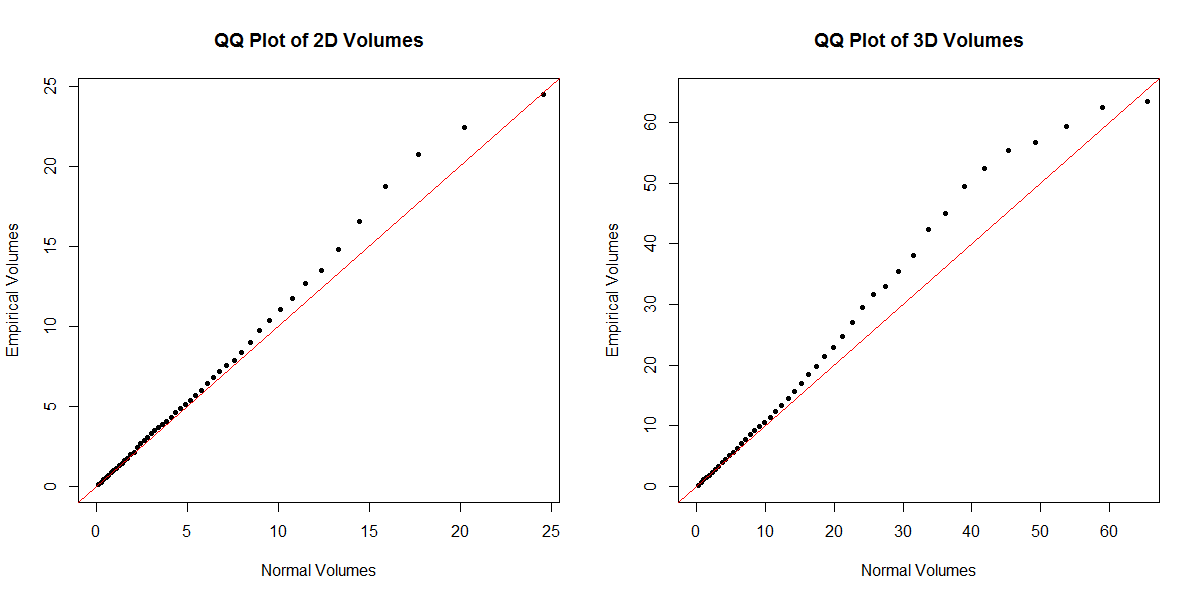}
\caption{QQ plot of $\widehat{V}_n (i/50)$ ($i=1,\ldots,49$) from a two-dimensional (left) and a three-dimensional (right) spherical normal sample of size~$n~\!=~\!1000$ against the theoretical volumes. The red line represents the identity function.  }
\label{multivariate_normal_gof}
\end{figure}

\section{Analysis of Multivariate Regularly Varying Distributions} 
Multivariate extreme value modelling has become the basic methodology in modelling and analyzing extreme multivariate risk. The use of  max-stable distributions and distributions in the domain of attraction of max-stable distributions as discussed in Beirlant et al. (2004) and de Haan and Ferreira (2006), translated into extreme value copulas, are fundamental for this purpose. Here we restrict to the case of multivariate regularly varying distributions, for which the underlying distribution ${\rm P}_{\bf X}$ of ${\bf X}$ is in the domain of attraction of a max-stable distribution. This class constitutes a direct generalization of the univariate Pareto-type distributions with distribution function given by $1-F(x)=x^{-1/\gamma}\ell (x)$ with $\gamma>0$ denoting the  {\it extreme value index} (EVI) measuring the tail heaviness and $\ell$ denoting a slowly varying function at infinity:
$$\lim_{t\to\infty}
{\ell (tx) \over \ell (t)} = 1\quad \mbox{ for every } x>0.
$$
See Cai et al. (2015) for a recent reference on bivariate risk estimation under the regular variation model. de Valk and Segers (2019) provide general results concerning the tail limits of optimal transports for regularly varying probability measures. 
Whereas the estimation of $\gamma >0$ in the univariate case has known an explosion of references starting with Hill (1975), the estimation of the EVI   in the multivariate case has recently taken off with Dematteo and Cl\'emen\c{c}on (2016) and Kim and Lee (2017). Here we provide an alternative approach  based on the constructed transports.

In the multivariate regular variation model, one assumes that there exists a   constant~$\gamma >0$ and a random unit vector ${\boldsymbol \Theta}$ with distribution ${\rm P}_\Theta$ over $\mathbb{S}_{d-1}$ such that, for all $x>1$, 
\begin{equation}
 {\rm P}_{\bf X}\left( \| \X \|/t > x, \, \X /\|\X\| \in \cdot \; | \; \| \X \| > t \right)
\longrightarrow_v x^{-1/\gamma}{\rm P}_\Theta({\boldsymbol \Theta} \in \cdot )
\label{mrv}
\end{equation}
as $t \to \infty$,   with $\to_v$ indicating vague convergence.  ${\rm P}_\Theta$ is  known as the {\it spectral measure}, while  the EVI $\gamma$ is also referred to as  the  index of regular variation of~${\rm P}_{\bf X}$.  That index is an important indicator of tail heaviness:   the higher  $\gamma$, the higher the spread in the  outer    tails;  $\gamma$ provides an evaluation of tail heaviness    in the different directions of~$\mathbb{R}^d$. 
For more details concerning spectral measure representations and spectral densities we refer to Chapter 8 in Beirlant et al.~(2004).
\\
A  relevant family of distributions satisfying condition \eqref{mrv} is given  by the vectors $\X$ for which the radii $R_{\X}= \| \X \|$ follow a Pareto-type distribution with 
$${\rm P}_{\bf X}
(R_{\X} > r \,| \,  \X /\|\X\|) \sim C_{\X /\|\X\|} \, r^{-1/\gamma}\quad \text{ as } r \to \infty$$
 with the scale parameter $C$ depending on the direction $\X /\|\X\|$ through the angular distribution of  ${\boldsymbol \Theta}$ on $\mathbb{S}_{d-1}$, while $\gamma$ does not depend on   ${\X /\|\X\|}$.

We propose to evaluate the risk associated with ${\rm P}_{\bf X}$ satisfying~\eqref{mrv} by constructing Pareto QQ plots based on  ${\bf Q}_{n}$, $\widetilde{\bf Q}_{n}$ or  $\widehat{\bf Q}_{n, \xi}$. More precisely, we discuss here the random variables 
$$\widehat{Y}_i :=\langle {\bf u}_{i}, {\bf Q}_{n} ({\bf u}_i)\rangle = \langle {\bf F}_{n} ({\bf X}_i), {\bf X}_i \rangle  , \quad i=1,2,\dots,n.$$ In the case studies when putting the link with  the risk surfaces and $\hat{\rho}^{\bf X}_n$ risk measures we use the versions $\langle {\bf u}_{i}, \widehat{\bf Q}_{n, \xi} ({\bf u}_i)\rangle$ based on $\widehat{\bf Q}_{n, \xi}$ rather than ${\bf Q}_{n}$.
Denote by $$\widehat{Y}_{1,n}<\widehat{Y}_{2,n}< \dots < \widehat{Y}_{n,n}$$ the corresponding  order statistics and by $$Y_i := \langle {\bf F}_\pm ({\bf X}_i), {\bf X}_i\rangle = \langle {\bf U}_i, {\bf Q}_\pm({\bf U}_i)\rangle$$ the population version of   $\widehat{Y}_i$:    here,~${\bf U}_i= {\bf F}_\pm ({\bf X}_i)$, $i=1,\ldots, n$ are i.i.d.\ with uniform distribution ${\rm U}_d$. 

\vspace{0.3cm}
Considering the particular case of an elliptical   regularly varying distribution, let $\X$ be such that 
$$\boldsymbol{\Sigma}^{-1/2} ({\bf X}- {\boldsymbol \mu})=
{{\bf U} \over \| {\bf U} \|} Q_R (\| {\bf U} \| )\quad\text{ with }\quad  {\bf U}\sim{\rm U}_d,$$
and $Q_R (p) = (1-p)^{-\gamma}\ell_Q(1/(1-p))$ where  $\ell_Q$ a slowly varying function, i.e.
and $\boldsymbol{\Sigma}$ has full rank. Then, $Y_i=_dY$ where, in view of \eqref{Qell},
\begin{eqnarray}
Y
 &:=  & \langle {\bf U}, {\boldsymbol\Sigma}^{1/2}{{\bf U}\over \| {\bf U}\|}\rangle
 \; Q_R (\| {\bf U}\|) + \langle {\bf U},{\boldsymbol \mu} \rangle \nonumber \\
&=& \langle { {\bf U} \over \| {\bf U} \| }, {\boldsymbol\Sigma}^{1/2}{{\bf U}\over \| {\bf U}\|}\rangle
\| {\bf U}\| \; Q_R (\| {\bf U}\|) + \langle {\bf U},{\boldsymbol \mu} \rangle
\label{UX_ell}
\end{eqnarray}
and  
\begin{eqnarray*}
\log Y &= & -\gamma \log (1- \| {\bf U} \|)
 + \log \ell_Q ( 1/ (1-\| {\bf U} \|))+ \log \| {\bf U}\| + \log \langle {\bf U}, {\boldsymbol \mu} \rangle  \\
 && \hspace{0.5cm}+
 \log \langle { {\bf U} \over \| {\bf U} \| }, {\boldsymbol\Sigma}^{1/2}{{\bf U}\over \| {\bf U}\|}\rangle .
 \end{eqnarray*}
Let $V_u(z) = 1-e^{-x}/(1-u)I_{[z>\vert\log (1-u)\vert]}$ denote the shifted exponential  distribution function   with location $u$ and  quantile function 
 $Q_u(p)= -\log (1-p) - \log (1-u)$,  $p \in (0,1)$. 
Elementary algebra shows  that the distribution of $\log Y$ in \eqref{UX_ell}  conditional  on~$\| {\bf U}\|>u $ is such that 
$$ \mathrm{P}_{\X}\big(\log Y \leq y\vert\, \| {\bf U}\|>u
\big)=  \gamma V_u  \left( 1+ o(1) \right)\quad\text{as $u \to 1$}$$
since slowly varying functions satisfy $\log \ell_Q (x)/\log x \to 0$ as $x \to \infty$ and~$ 
{\langle {\bf U}, {\boldsymbol\Sigma}^{1/2}{\bf U}\rangle
/ \| {\bf U}\|^2}
$ is bounded from below and from above by the smallest and largest eigenvalues of ${\boldsymbol\Sigma}^{1/2}$, respectively  (these eigenvalues are strictly positive since  ${\boldsymbol\Sigma}$ is positive definite). Note that the $o(1)$ term linked to $\log \langle { {\bf U} \over \| {\bf U} \| }, {\boldsymbol\Sigma}^{1/2}{{\bf U}\over \| {\bf U}\|}\rangle$ depends on the direction of ${\bf U}$. 
 Using  Proposition \ref{weakconv} below concerning  the weak convergence of the empirical distribution based on the $\widehat{Y}_i$ rv's to the distribution of the $Y_i$ ($i=1,\ldots,n$), the Pareto QQ plot (plotting the ordered $\log \widehat{Y}$ data against the expected values of  standard exponential order statistics)
 \begin{equation}
\left( -\log (1- {n-j+1 \over n+1}),\log {\widehat Y}_{n-j+1,n}\right),\quad j=1,\ldots,n,
\label{Pa_QQ}
\end{equation}
   is  expected  
to exhibit  a linear pattern with slope $\gamma$  for large values of $\log \widehat{Y}$ when restricting to the points (with $j=1,\ldots,k$)	lying above the log-threshold point $$\left( \log {n+1 \over k+1},\log {\widehat Y}_{n-k,n}\right), \;\; k \in \{2,\ldots,n-1\}
$$ with $k/n$ small enough.

\begin{figure}[htbp]
\centering
\includegraphics[width=150mm]{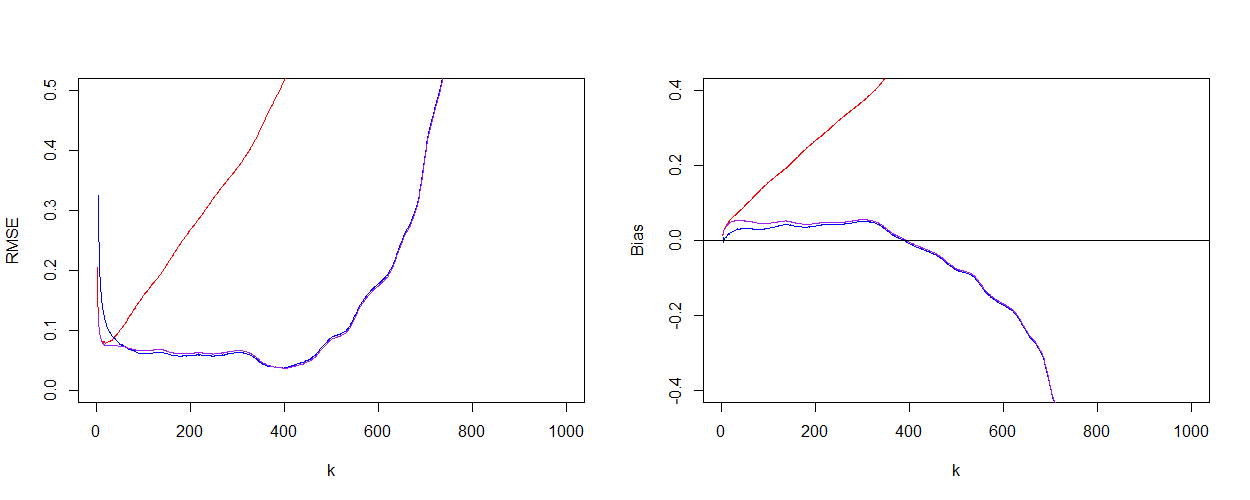}
\includegraphics[width=150mm]{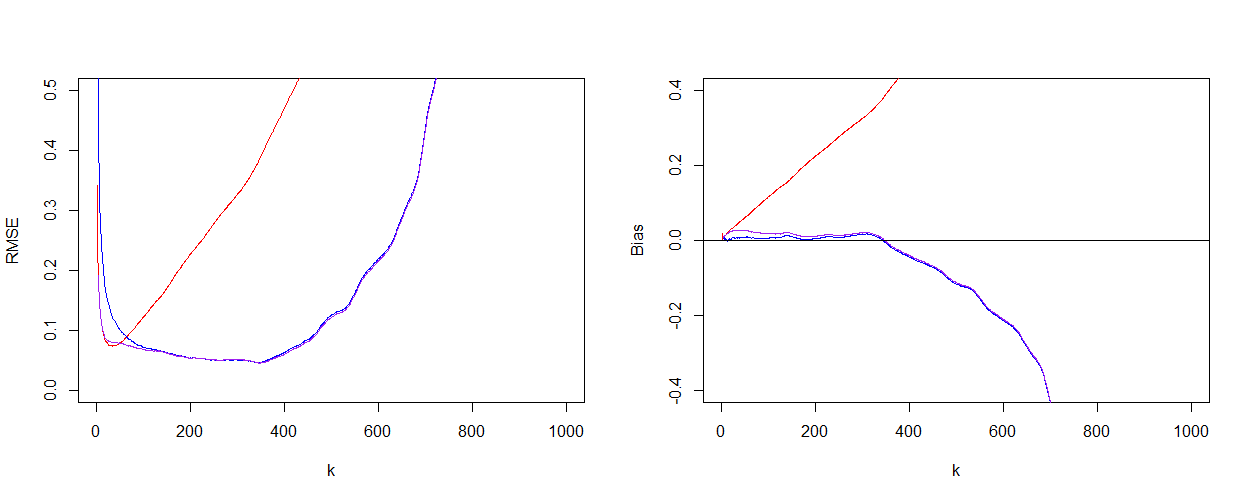}
\includegraphics[width=150mm]{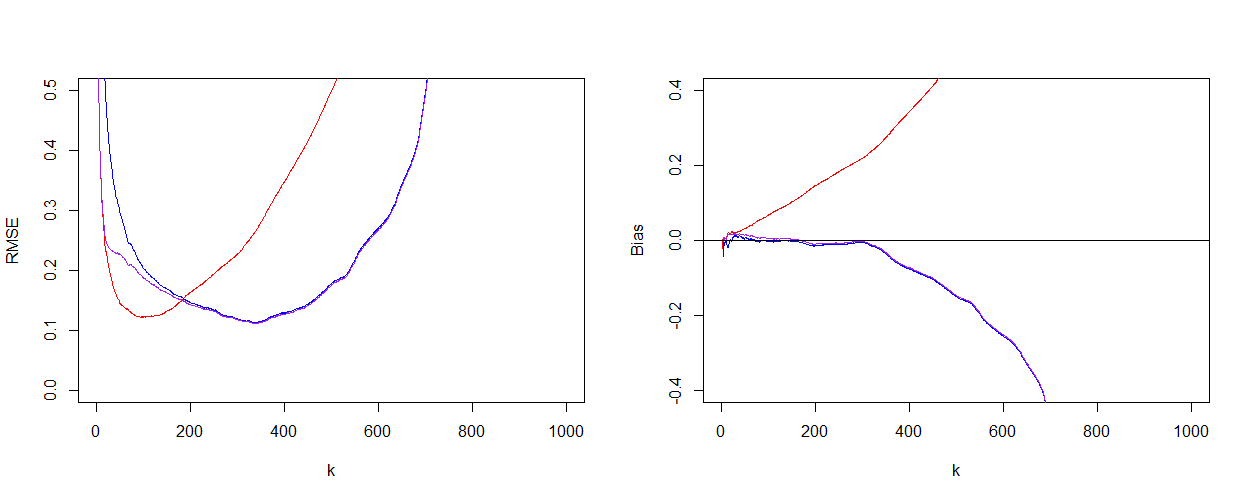}
\caption{Bivariate elliptical hyperbolic distribution with an EVI of $1/5$ (top), $1/3$ (center) and $1$ (bottom): RMSE (left) and bias (right) plots of the Hill (red), least-squares (blue) and ridge regression (purple) estimates for the EVI $\gamma$.}
\end{figure}

\vspace{0.3cm}
In the general multivariate regular variation case, \eqref{mrv} states that when restricting to radii above a certain threshold $t$, the transport of the exceedances ${\bf X}/t$ to ${\bf U}$  basically can be decoupled into a transport of the angular measure ${\boldsymbol \Theta}$ to the uniform distribution on~$\mathcal{S}_{d-1}$ and the probability integral transform of the radii to the uniform $\mathrm{U}_{[0,1]}$ distribution, to be compared with \eqref{Fell}. From this also an ultimate linear pattern of the Pareto QQ plot \eqref{Pa_QQ} can be expected under regularity conditions on the spectral measure of $\boldsymbol \Theta$. This will be pursued in a future work.


The slope of the QQ plot \eqref{Pa_QQ} at the top $k$ points  can be measured using the classical approach initiated in Hill~(1975),  by the average vertical increase $\log \widehat{Y}_{n-k,n}$  in the above Pareto QQ plot  :   
\begin{align*}
H_{k,n} = {1 \over k }\sum_{j=1}^k \log \frac{ \widehat{Y}_{n-j+1, n}}{\widehat{Y}_{n-k, n}} = \sum_{j=1}^k 
j \,\log \frac{ \widehat{Y}_{n-j+1, n}}{\widehat{Y}_{n-j, n}}, \;\; k= 1 ,\ldots,n.
\end{align*}
Alternatively, bias reduction methods are available, such as the ridge regression method proposed by  Buitendag et al.~(2019):
\begin{align}
\hat{\gamma}_{k}(\tau) & = H_{k,n}- \frac{ \bar{c}_k  \,  \sum_{j=1}^k (c_{j,k}-\bar{c}_k)\, j \,\log \frac{ \hat{Y}_{n-j+1, n}}{\hat{Y}_{n-j, n}} }{ \sum_{j=1}^k (c_{j,k}-\bar{c}_k)^2+ k\, \tau},
\end{align}
with $c_{j,k} = \left( {j \over k+1} \right)^{-\rho}$ where $\rho<0$.
The choice $\tau=0$ of the ridge parameter $\tau$ leads to simple least squares regression $\hat{\gamma}_{k}^{ls}$ as introduced in  Feuerverger and Hall (1999) and Beirlant et al.~(1999) for univariate Pareto-type tail estimation.


 In Figures 6 to 9,  simulation results are reported showing that the proposed methods based on the $\widehat{Y}_i$ variables $( i=1,\ldots,n)$, provides a promising   estimator of~$\gamma$. For bivariate and trivariate elliptical hyperbolic and  $t$ distributions and sample size $n=1000$ we report the bias and RMSE  as a function of $k$  for $H_{k,n}$, $\hat{\gamma}_{k}^{\text{ls}} =\hat{\gamma}_{k}(0) $ and $\hat{\gamma}_{k}^{\text{ridge}} =\hat{\gamma}_{k} \left( \hat{\tau}_k \right)$, with $\hat{\tau}_k$  a data-driven choice of the ridge parameter  as proposed in Buitendag et al.~(2019).   These results compare favourably with the simulation results in Figure 1 in Kim and Lee (2017).   Note also that the computational effort when computing the $\widehat{Y}$ values is less demanding in comparison with the volume computations in the preceding section.  

\vspace{0.3cm} We end this section stating the weak convergence of the empirical distribution  based on $\widehat{Y}_i$ ($i=1,\ldots,n$) to the $Y$ distribution. To this end let $\widehat{G}_n$ denote the empirical distribution function of the $\widehat{Y}_i$, and $G$ the distribution function of $Y$.
\begin{proposition}\label{weakconv}
Assuming $\mathbb{E}(\| {\bf X}\|) < \infty$, then 
$$
\widehat{G}_n \to_d G, \mbox{ as }  n \to \infty.
$$
\end{proposition}
\noindent\textbf{Proof.}   Following Theorem 11.3.3(b) in Dudley (2004), it suffices to prove that as $n \to \infty$ 
$$
\int g(x)d \widehat{G}_n (x)  = {1 \over n}\sum_{i=1}^n g(\widehat{Y}_i) \to \int g(x) dG(x)
$$
for any bounded Lipschitz function.

First, since $Y_1,\ldots, Y_n$ are i.i.d.,  for any bounded continuous function $g$ we have that as $n \to \infty$ 
$$
{1 \over n}\sum_{i=1}^n g(Y_i) \to  \int g(x) dG(x).
$$
Also  if $g$ is  Lipschitz with constant $M$ 
\begin{eqnarray*}
| {1 \over n}\sum_{i=1}^n g(\widehat{Y}_i) - {1 \over n}\sum_{i=1}^n g(Y_i) | &\leq &  
M \;{1 \over n}\sum_{i=1}^n |\widehat{Y}_i-Y_i|\\
&\leq & M \|{\bf F}_n-{\bf F}_{\pms} \|_{\infty} \; \left( {1 \over n}\sum_{i=1} \|{\bf X}_i \|\right), 
\end{eqnarray*}
which tends to 0 a.s. if $\mathbb{E}(\| {\bf X}\|) < \infty$.
\qed\medskip

\begin{figure}[htbp]
\centering
\includegraphics[width=150mm]{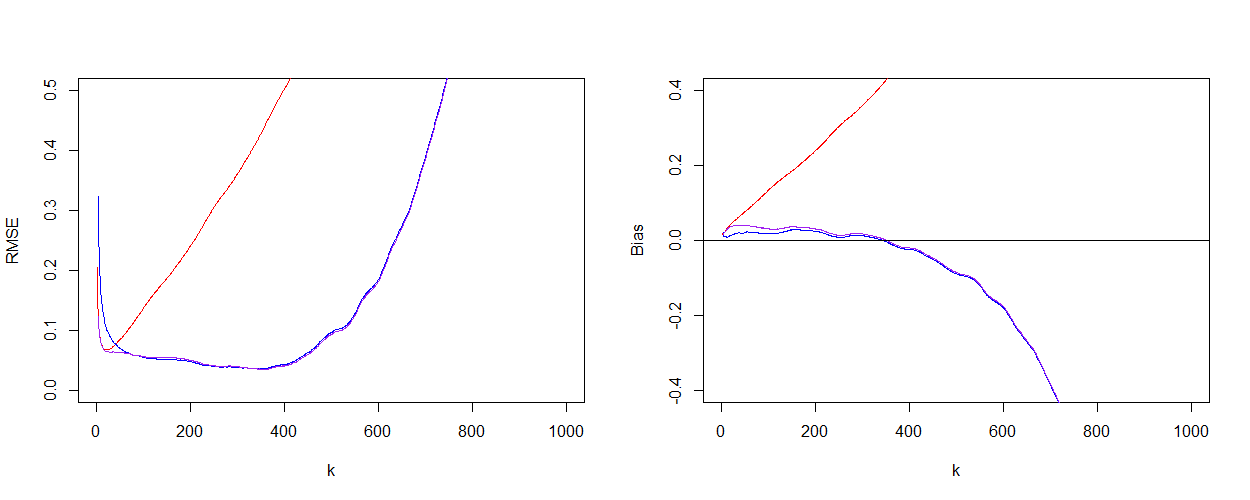}
\includegraphics[width=150mm]{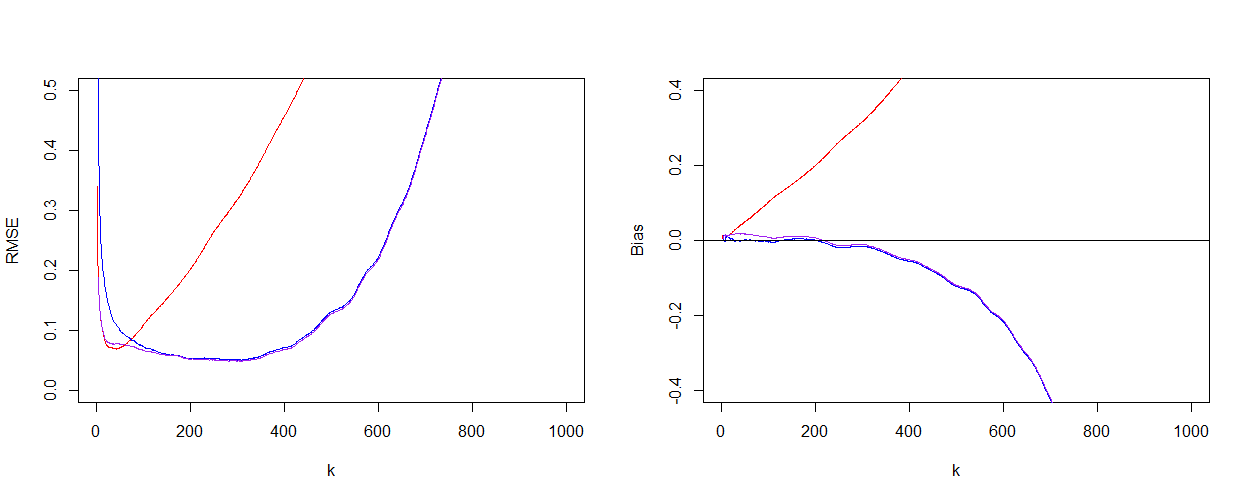}
\includegraphics[width=150mm]{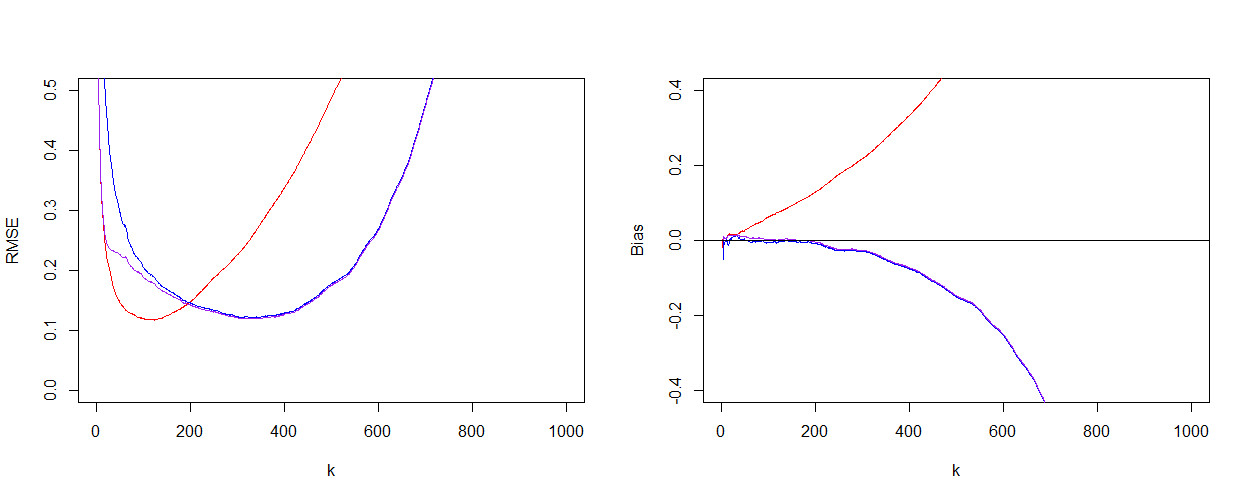}
\caption{Bivariate Student-t distribution with an EVI of $1/5$ (top),~$1/3$ (center) and $1$ (bottom): RMSE (left) and bias (right) plots of the Hill (red), least-squares (blue) and ridge regression (purple) estimates for the~EVI $\gamma$.}
\end{figure}

\begin{figure}[htbp]
\centering
\includegraphics[width=150mm]{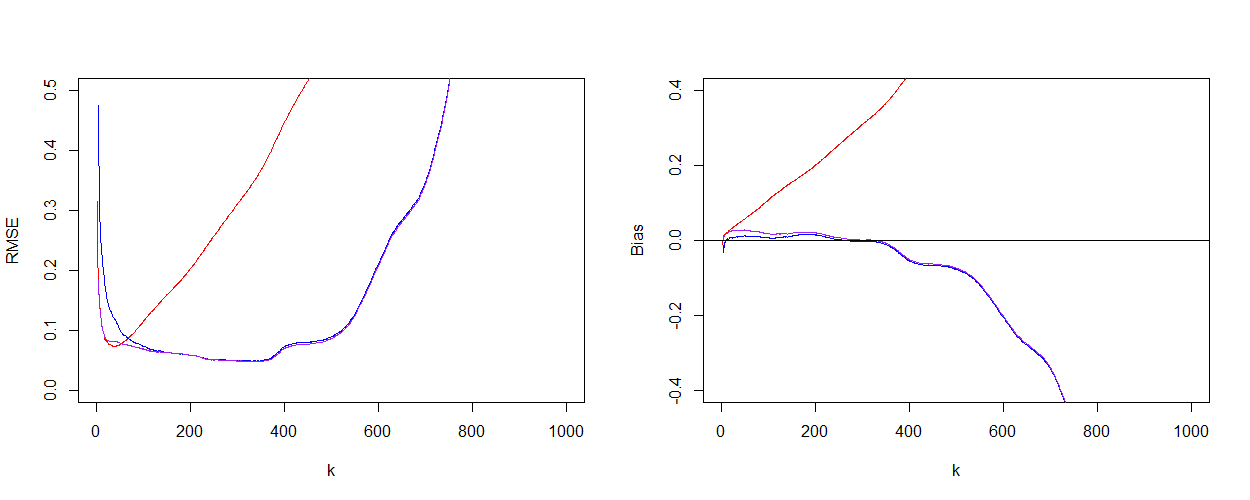}
\includegraphics[width=150mm]{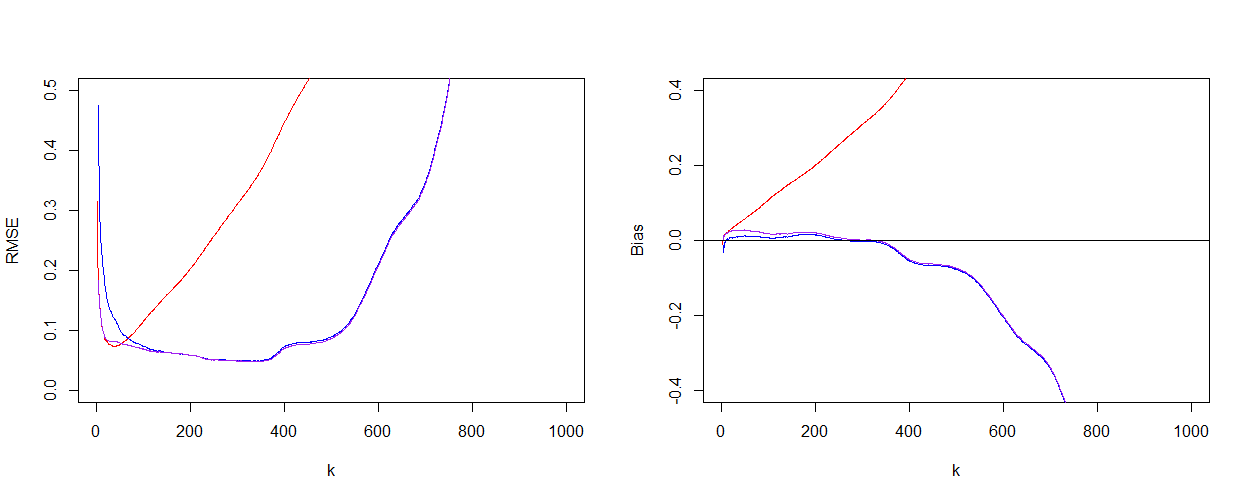}
\includegraphics[width=150mm]{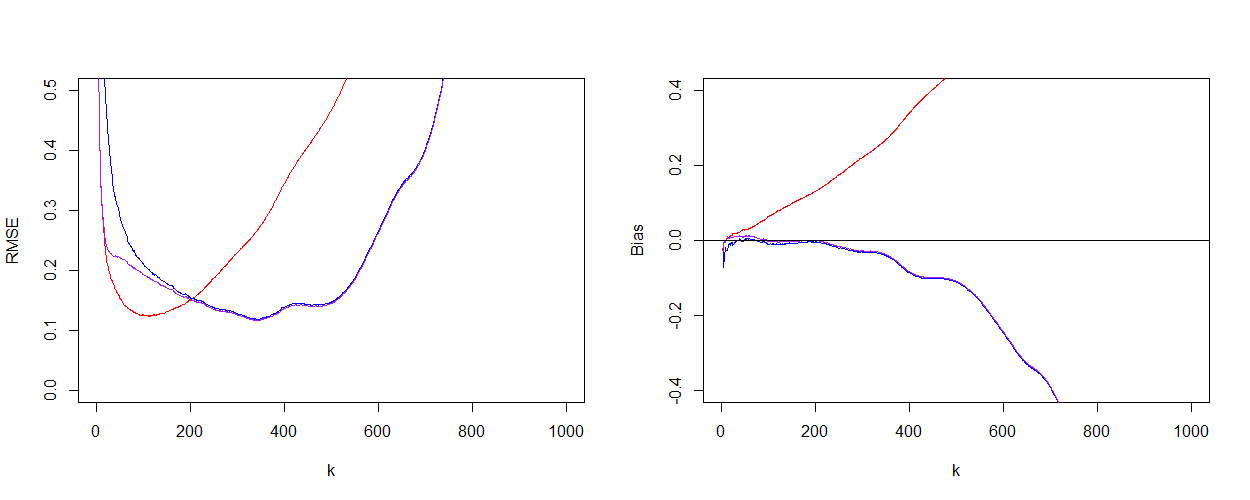}
\caption{3 dimensional elliptical hyperbolic distribution with an EVI of~$1/5$ (top), $1/3$ (center) and $1$ bottom: RMSE (left) and bias (right) plots of the Hill (red), least-squares (blue) and ridge regression (purple) estimates for the~EVI $\gamma$. }
\end{figure}

\begin{figure}[htbp]
\centering
\includegraphics[width=150mm]{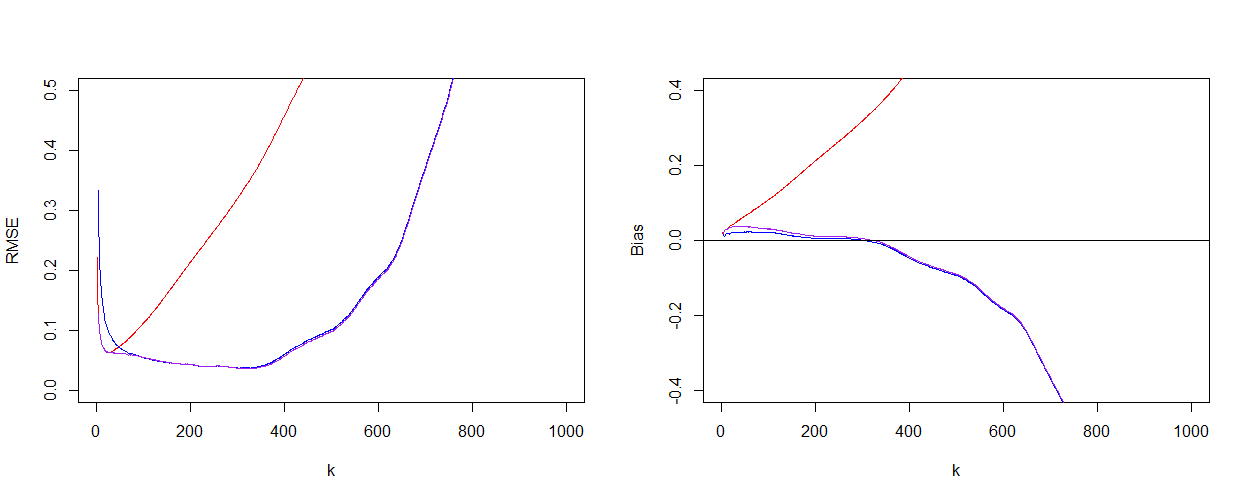}
\includegraphics[width=150mm]{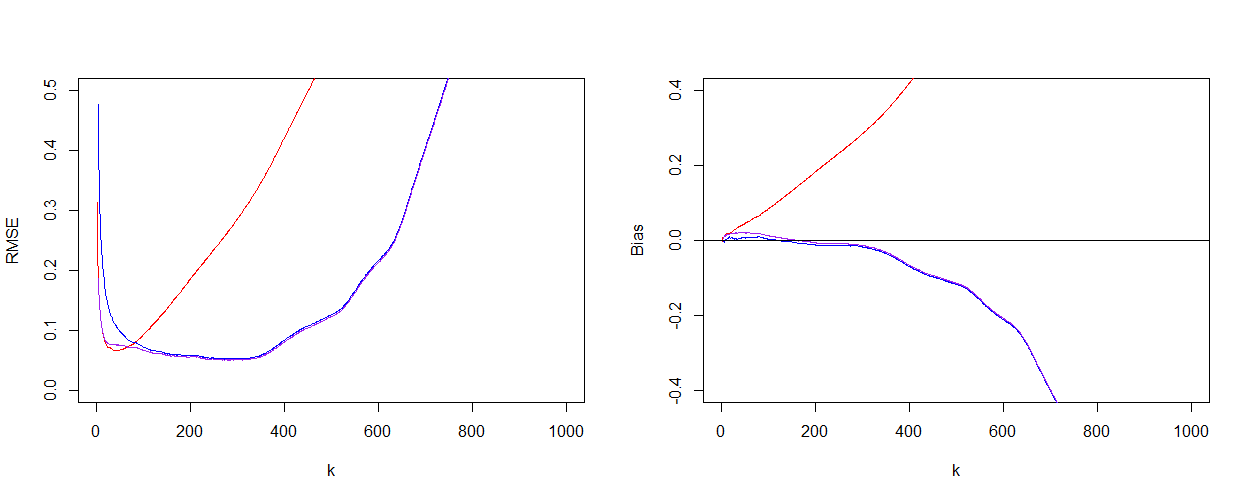}
\includegraphics[width=150mm]{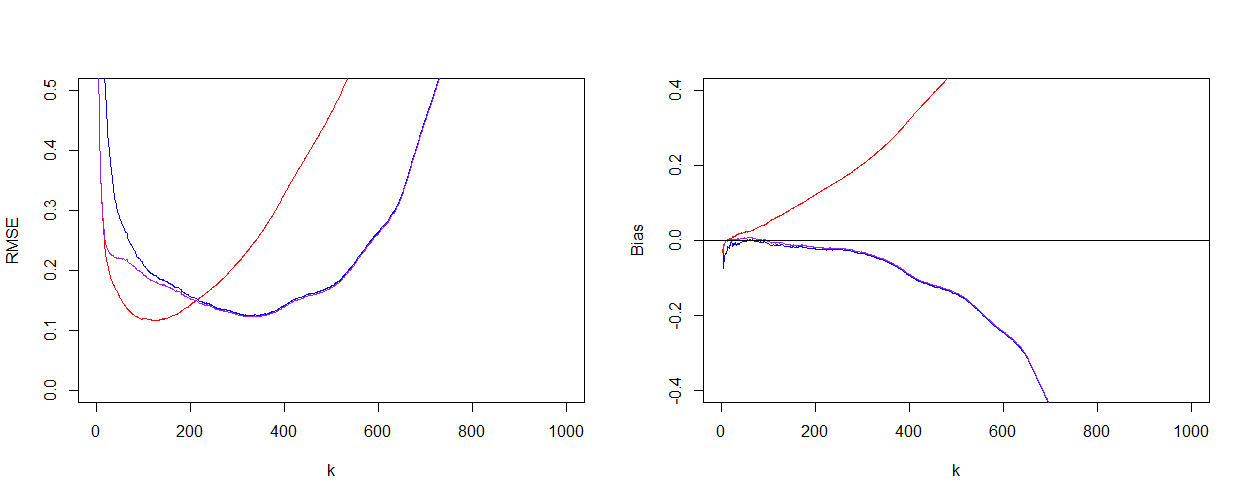}
\caption{3 dimensional Student-t distribution with an EVI of $1/5$ (top),~$1/3$ (center) and $1$ (bottom): RMSE (left) and bias (right) plots of the Hill (red), least-squares (blue) and ridge regression (purple) estimates for the~EVI $\gamma$.}
\end{figure}

\section{Case studies}

\subsection{Google Apple share log-returns} 

As a first case study we consider the daily log-returns of Google and Apple share prices for the period 01-01-2005 to 01-04-2019. 
We group three years' data at the end of the final month and denote these as 12-2007, 01-2008, up to 03-2019. For instance, the 12-2007 group consists of the data from 01-01-2005 to 31-12-2007, the 01-2008 group consists of the data from 01-02-2005 to 31-01-2008, and so forth. In Figure 10 however time is shifted monthly.

  Clearly when the 2008 financial crisis period has disappeared from the  moving window, the risk surfaces given in Figure 11 become  less high, which is captured in  the risk measures $\hat{\rho}^{\bf X}_n$ and $\hat{\rho}^{\bf X}_{n, 0.1}$ in Figure 10 that reach a lower level as time progresses. From the Pareto QQ plots \eqref{Pa_QQ}  and the corresponding EVI estimates we observe that the slopes in the QQ plots at the highest $\log \widehat{Y}$ values (indicated in Figure 13 with regression fits to the right of  the  vertical line  at $k= 10$) do not really change for the three time windows considered. Indeed the slope estimate plots in Figure 12 indicate an EVI value  around~0.3 for the three time windows. Note however that the three Pareto QQ plots move downwards  with time, indicating lower $\log \widehat{Y}$ quantile values as time progresses. So we can conclude that the decrease in risk is not due to a change in EVI but rather in the scale of the observed Pareto-type behaviour. This is further illustrated in Figure 14 indicating in red the ten observations corresponding to the  highest $\widehat{Y}$ values of the Pareto QQ plots in the corresponding scatterplots. Note that for the time window 12-2009 both Google and Apple suffer a loss between 10 and 20\%, while for 12-2013 Google registered losses between 5 and 10 \% while Apple losses were smaller than 5 \%. Finally, in the final time window, the worst losses for Apple are between 6 and 7 \% and between 0 and 6\%. This allows for a more specific interpretation to the risk increase.   

\begin{figure}[ht!]
\centering
\includegraphics[width=150mm]{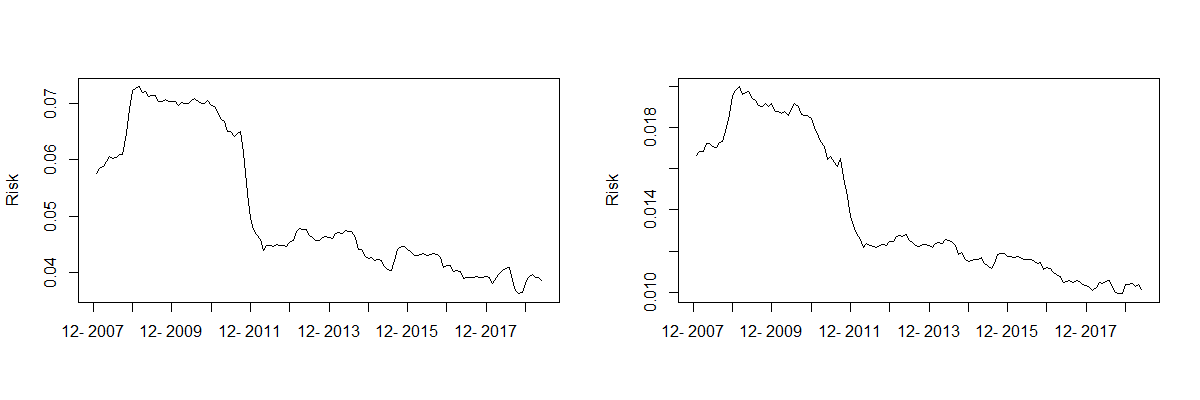}
\caption{The multivariate risk measures  $\hat \rho^{\bf X}_{n}$ (left) and $ \hat \rho^{\bf X}_{n,0.1}$ (right) for the Google and Apple log-returns data for monthly periods from 2007 to 2019.  }
\end{figure}

\begin{figure}[ht!]
\centering
\includegraphics[width=50mm,trim=1cm 2cm 1cm 0cm,clip]{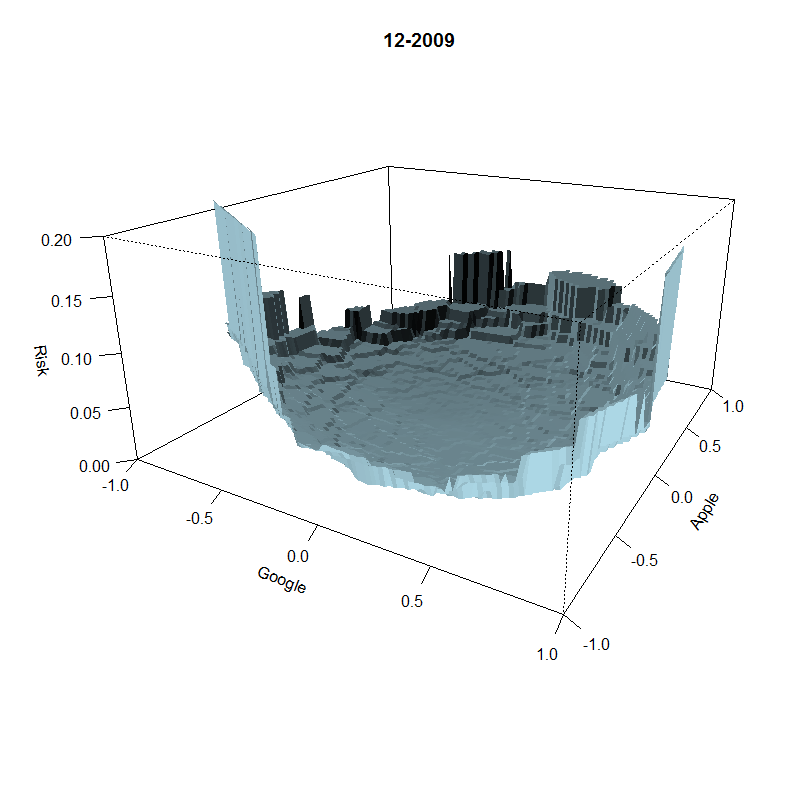}
\includegraphics[width=50mm,trim=1cm 2cm 1cm 0cm,clip]{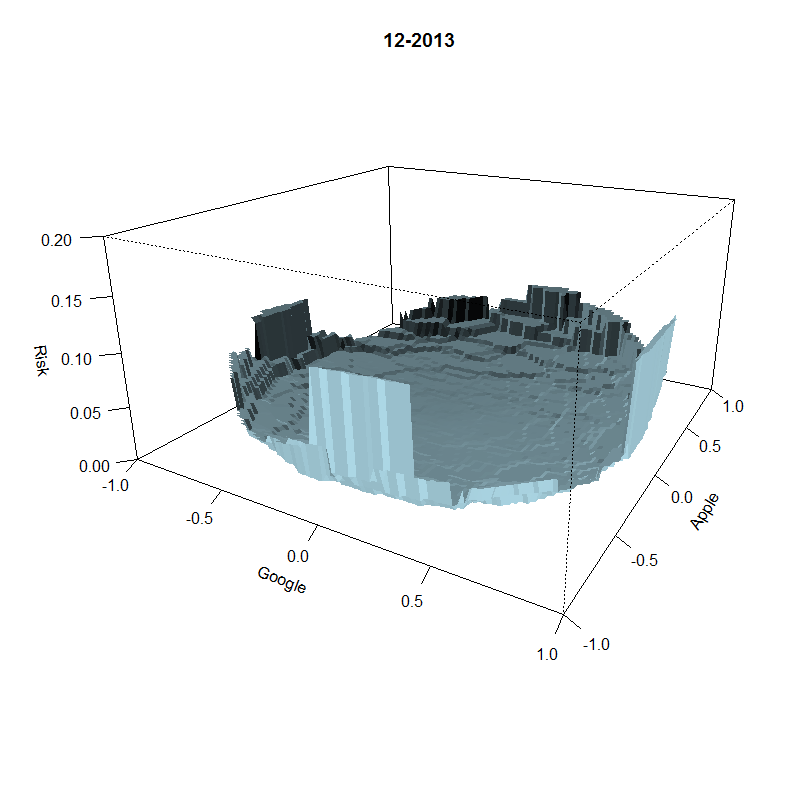}
\includegraphics[width=50mm,trim=1cm 2cm 1cm 0cm,clip]{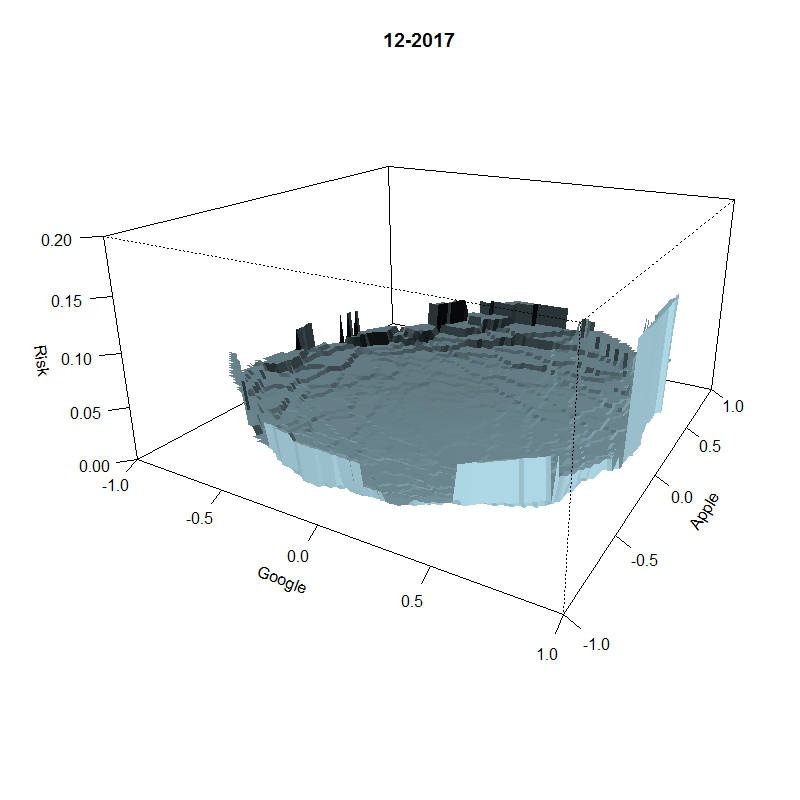}
\caption{The bivariate risk surfaces for the Google and Apple log-returns data for the datasets 12-2009, 12-2013 and 12-2017.  }
\end{figure}

\begin{figure}[ht!]
\centering
\includegraphics[width=150mm]{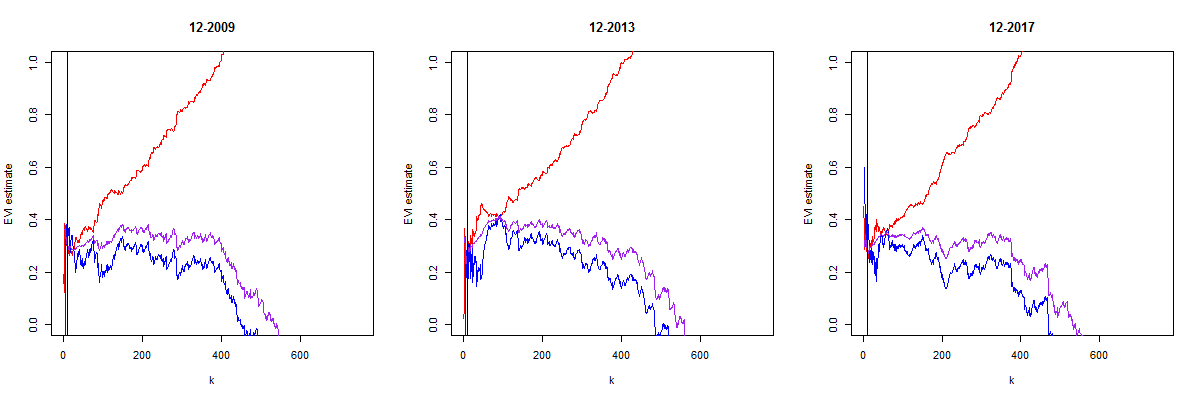}
\caption{EVI estimates for the Google and Apple log-returns datasets 12-2009 (left), 12-2013 (center) and 12-2017 (right). The red, purple and blue lines represent the Hill, ridge regression and least squares EVI estimators respectively. }
\end{figure}

\begin{figure}[ht!]
\centering
\includegraphics[width=130mm]{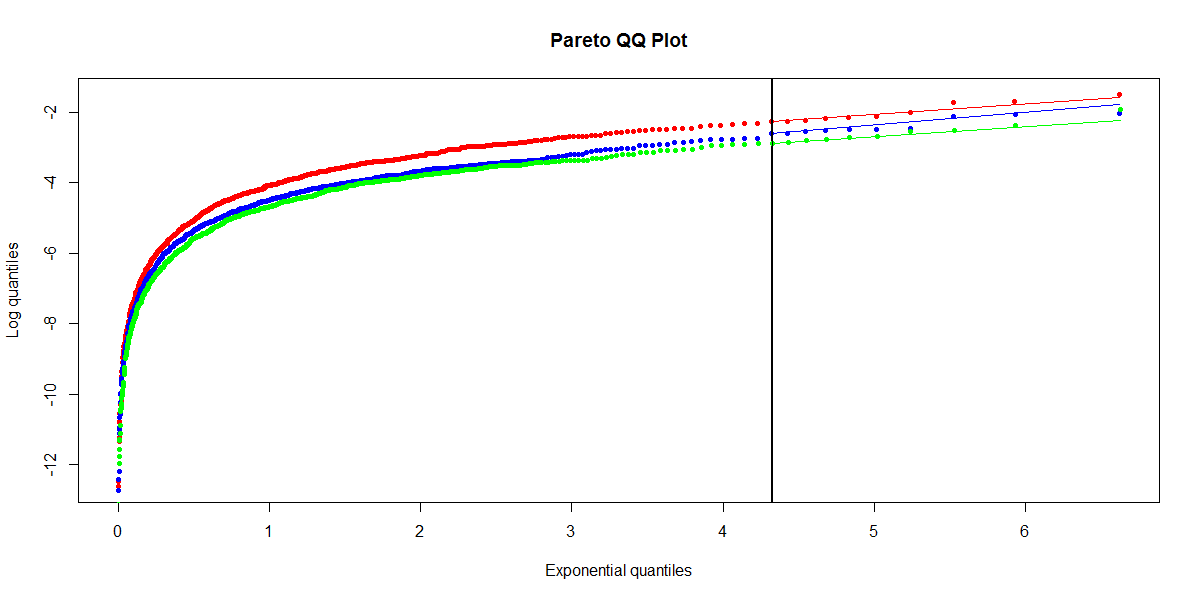}
\caption{Pareto QQ plots of  $\langle {\bf u}_i, \widehat{\bf Q}_{n,\xi} ({\bf u}_i) \rangle$ from the Google and Apple log-returns datasets 12-2009 (red), 12-2013 (blue) and 12-2017 (green)}
\end{figure}

\begin{figure}[ht!]
\centering
\includegraphics[width=100mm]{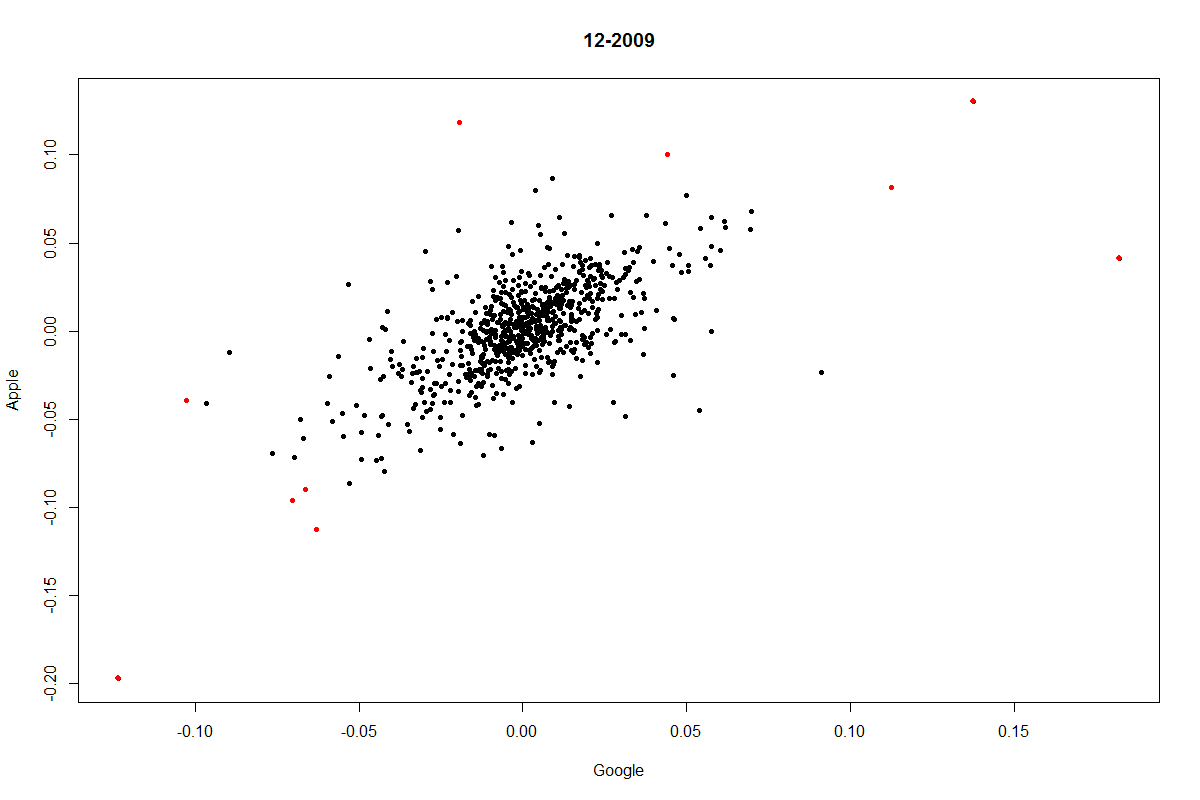} \\
\includegraphics[width=100mm]{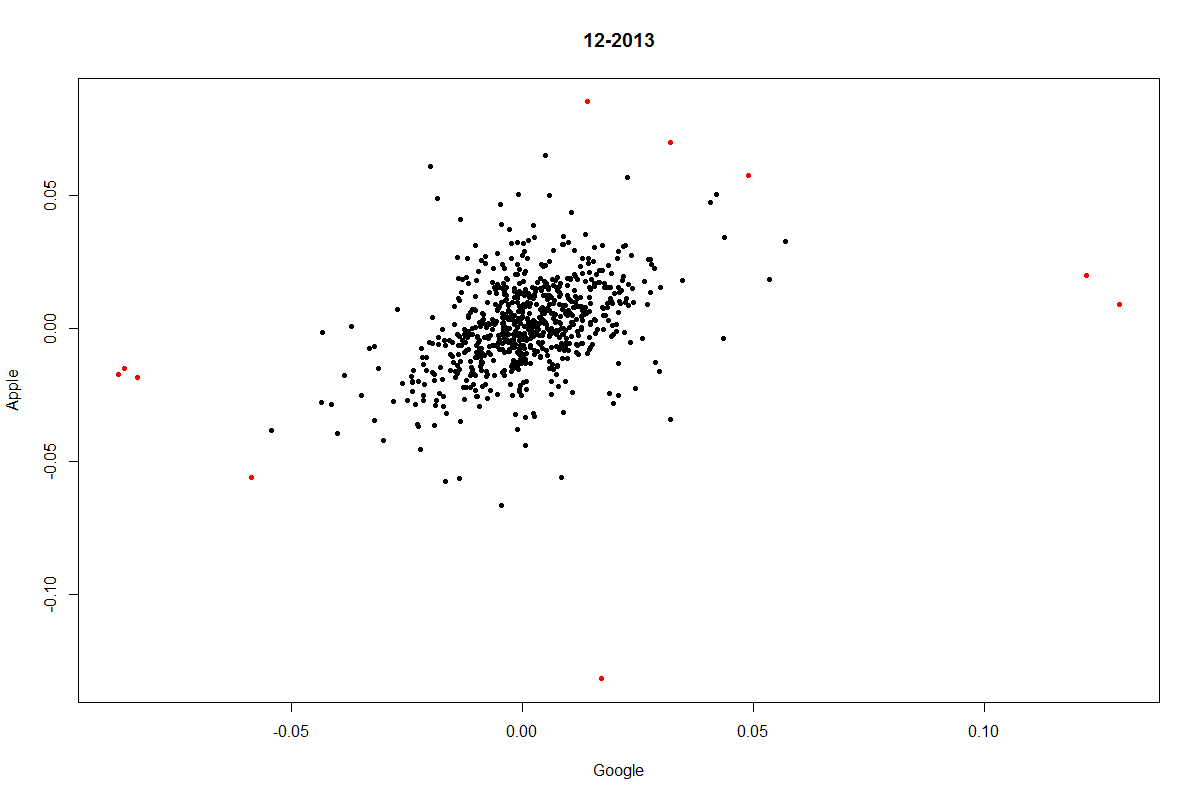}\\ \includegraphics[width=100mm]{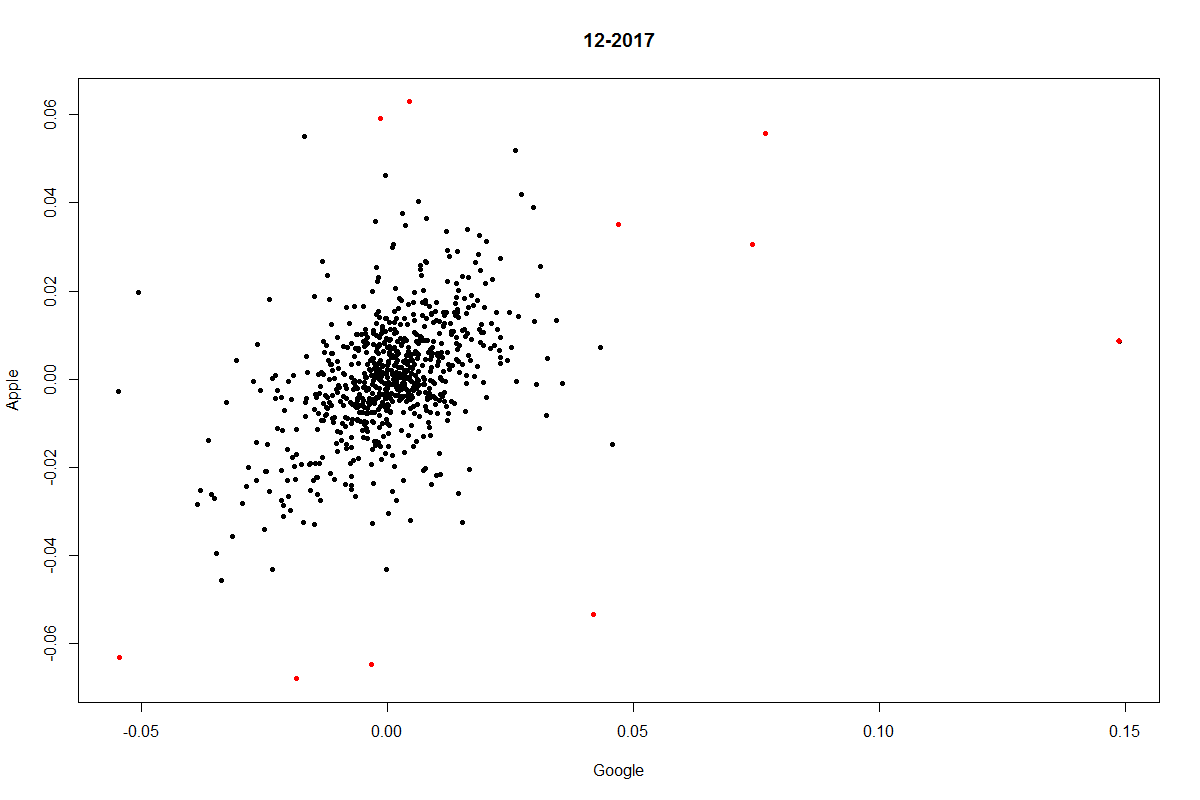}
\caption{Scatterplots  of  the Google and Apple log-returns data sets 12-2009 (top), 12-2013 (middle) and 12-2017 (bottom) indicating the ${\bf X}$ observations corresponding to the ten highest $\widehat{Y}$ values.\bigskip\bigskip\bigskip\bigskip\bigskip}
\end{figure}

\subsection{Danish fire insurance}

As a further case study we consider the log contents, building and profit amounts of the Danish fire insurance data for the period 1980 to 1990, discussed for instance in Embrechts et al.~(1997) or Beirlant et al.~(2004).
We again group three years' data at the end of the final month and denote these as 12-1982, 01-1983, up to 12-1990. For instance, the 12-1982 group consists of the data from 01-01-1980 to~31-12-1982, the 01-1983 group consists of the data from 01-02-1980 to 31-01-1983, and so forth. Here the risk increases starting at the time period 1985-1988, while the EVI stays constant at 0.2. Here the risk increases with time due to increasing levels of the highest $\widehat{Y}$ observations rather than due to an increase of the EVI. While the classical Danish fire insurance data set has been discussed in several actuarial statistics papers, this increase in (multivariate) risk has not been observed before.
This is further illustrated in Figure 18 indicating in red the ten observations corresponding to the  highest~$\widehat{Y}$ values of the Pareto QQ plots in the corresponding scatterplots (in the original scale). Note the increase in scales  for the second time window.  Moreover the largest risks are  mainly influenced by the  building and contents losses. Concerning the building-contents scatterplot the indicated losses in the first time period are concentrated at the combinations of (low building, high contents)  and (high building, middle to high contents) values of the (building, contents) vector. However, for the second time period, a shift towards the (high building, low contents) and (high building, high contents) values is observed.

\begin{figure}[ht!]
\centering
\includegraphics[width=150mm]{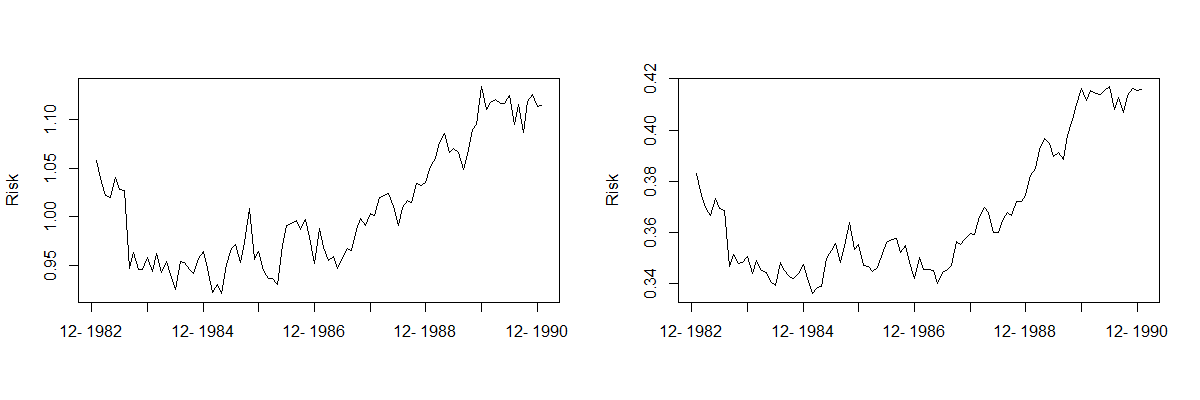}
\caption{The multivariate risk measures  $\hat \rho^{\bf X}_{n}$ (left) and $ \hat \rho^{\bf X}_{n,0.1}$ (right) for the Danish fire insurance data for monthly periods from 1980 to 1990. }
\end{figure}

\begin{figure}[ht!]
\centering
\includegraphics[width=150mm]{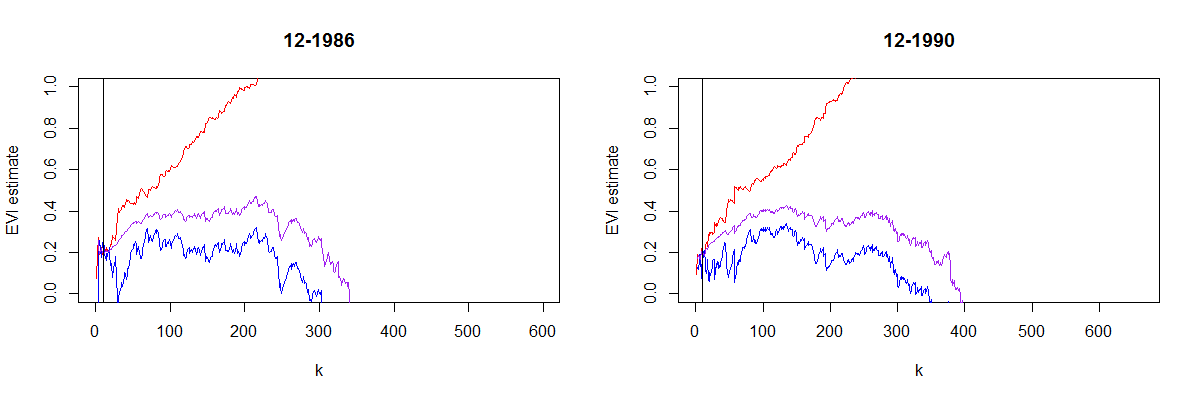}
\caption{Potential-based EVI estimates for the Danish fire insurance datasets 12-1986 (left) and 12-1990 (right). The red, purple and blue lines represent the Hill, ridge regression and least squares EVI estimators respectively.}
\end{figure}

\begin{figure}[ht!]
\centering
\includegraphics[width=130mm]{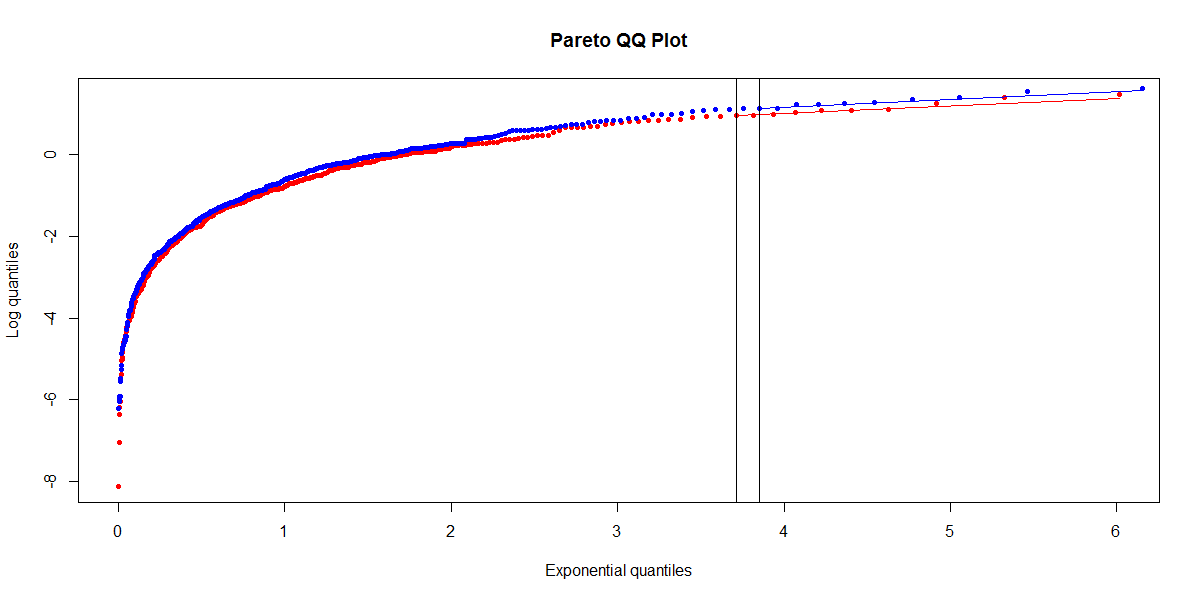}
\caption{Pareto QQ plots of the potential points $\langle {\bf u}_i, \widehat{\bf Q}_{n,\xi} ({\bf u}_i) \rangle$ from the Danish fire insurance datasets 12-1986 (red) and 12-1990 (blue)\bigskip\bigskip\bigskip}
\end{figure}

\begin{figure}[ht!]
\centering
\includegraphics[width=130mm]{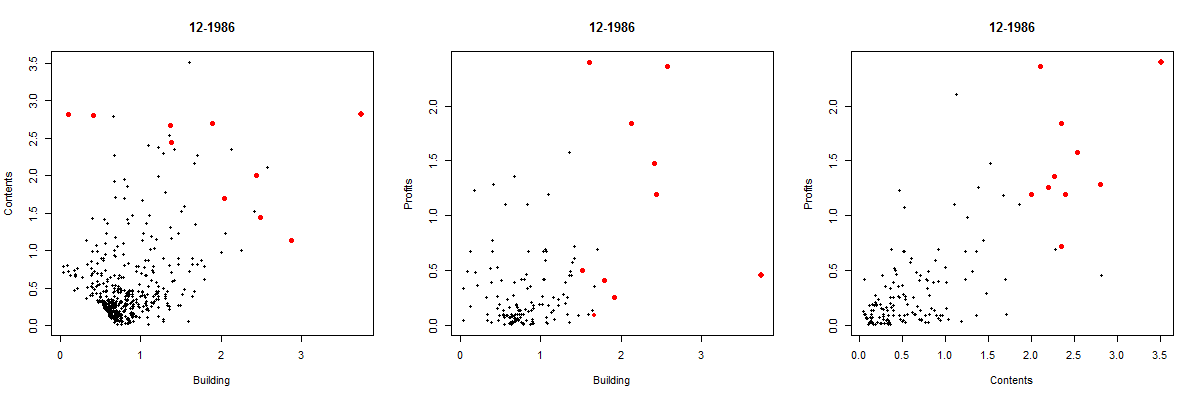} \\
\includegraphics[width=130mm]{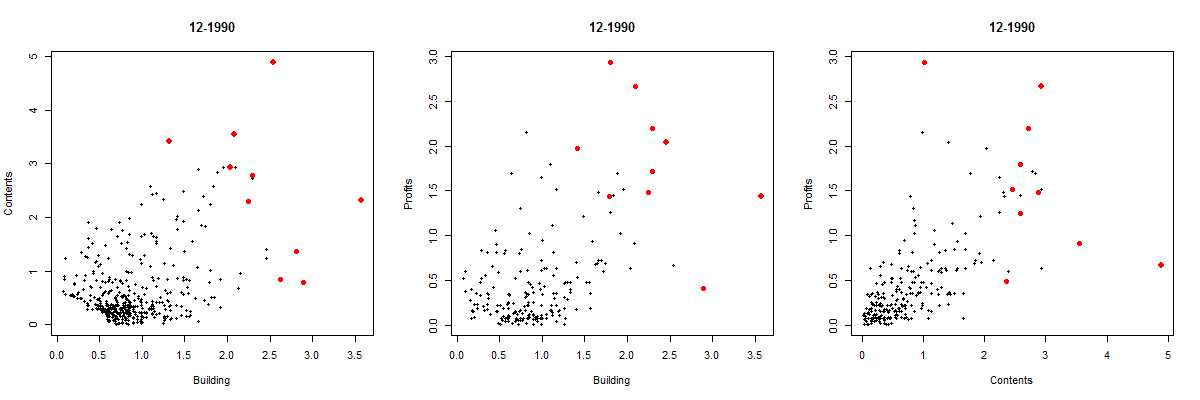}
\caption{Scatterplots  of  the Danish fire insurance data sets 12-1986 (top) and 12-1990 (bottom) indicating the ${\bf X}$ observations corresponding to the ten highest $\widehat{Y}$ values.\bigskip\bigskip}
\end{figure}

\section{Conclusion}

\noindent In this paper we illustrate the use of the recently developed concepts  of center-outward distribution and quantile functions as introduced in Hallin (2017) and del Barrio et al.~(2018) in multivariate risk measurement. Associated with   a new estimation method for the center-outward distribution and quantile functions, we indicate several new  multivariate risk measures, including a novel estimator of the index of multivariate regular variation which exhibits a good behavior in simulations. Clearly, subsequent work has to focus on establishing consistency and deriving the asymptotic distribution of the estimator of the index of multivariate regular variation.

\section*{Bibliography}

\noindent Beirlant, J.,  Dierckx, G.,  Matthys, G., and Y. Goegebeur, Y. (1999). Tail index estimation and an exponential regression model. {\it Extremes}, 5, 157-180. \\

\noindent Beirlant, J., Mason, D.M. and Vynckier, C. (1999). Goodness-of-fit analysis for multivariate normality based on generalized quantiles. {\it Computational Statistics and Data Analysis}, 30, 119-142. \\

\noindent Beirlant, J., Goegebeur, Y.,  Segers, J.,  and Teugels, J. L. (2004). {\it Statistics of Extremes: Theory and Applications}. Wiley.\\

\noindent
Buitendag,  S., Beirlant, J., and  de Wet, T. (2019). Ridge regression estimators for the extreme value index. {\it Extremes}, \url{https://doi.org/10.1007/s10687-018-0338-4}.
\\

\noindent Boyd, S. and Vandenberghe, L. (2004). {\it Convex Optimization}. Cambridge University Press.
\\

\noindent Cai, J., Einmahl, J., de Haan, L. and Zhou, C. (2017). Estimation of the marginal expected shortfall: the mean when a related variable is extreme. {\it J. Royal Statistical Society, Ser. B}, 77, 417-442. \\

\noindent
Charpentier, A. (2018). An introduction to multivariate and dynamic risk measures. \url{https://hal.archives-ouvertes.fr/hal-01831481} \\

\noindent Chernozhukov, V., Galichon, A., Hallin, M., and Henry, M. (2017). Monge-Kantorovich depth, quantiles, ranks, and signs. {\it Annals of Statistics}, 45, 223-256.\\

\noindent de Haan, L. and Ferreira, A. (2006). {\it Extreme Value Theory: an Introduction}. Springer Series in Operations Research and Financial Engineering. \\ 

\noindent del Barrio, E., Cuesta Albertos, J., Hallin, M., and Matran, C. (2018). Smooth cyclically monotone interpolation and empirical center-outward distribution functions.  Working Papers ECARES 2018-15, ULB, \url{https://ideas.repec.org/p/eca/wpaper/2013-271399.html}. \\

\noindent del Barrio, E. and Loubes, J.M. (2019). Central limit theorems for empirical transportation cost in general dimension.
{\it Annals of Probability}, 47, 926--951.\\

\noindent  de Valk, C. and Segers, J. (2019). Tails of optimal transport plans for regularly varying probability measures. arXiv:1811.12061v2. \\

\noindent Dematteo, A. and Cl\'emen\c{c}on, S. (2016). On tail index estimation based on multivariate data. {\it Journal of Nonparametric Statistics}, 28, 152-176.\\

\noindent Dudley, R.M. (2004). {\it Real Analysis and Probability}. Cambridge University Press.\\

\noindent Einmahl, J. and Mason, D.M. (1992) Generalized quantile processes. {\it Ann. Statist.}, 20, 1062-1078.\\

\noindent Ekeland, I., Galichon, A., and Henry, M. (2012). Comonotonic measures of multivariate risks. {\it Mathematical Finance}, 22, 109-132. \\

\noindent Embrechts, P., Kl\"uppelberg, C., and Mikosch, T. (1997). {\it Modelling Extremal Events}. Springer, Berlin.\\

\noindent Feuerverger, A. and  Hall, P. (1999). Estimating a tail exponent by modelling departure from a Pareto distribution. \emph{Annals of Statistics}, 27, 760--781. \\

\noindent Figalli, A. (2018). On the continuity of center-outward distribution and
quantile functions.  {\it Nonlinear Analysis: Theory, Methods \& Applications}, 177, 413--421.\\ 

\noindent Gushchin, A.A. and Borzykh, D.A. (2017). Integrated quantile functions: properties and applications. {\em Modern Stochastics: Theory and Applications}, 4, 285--314. \\

\noindent Hallin, M. (2017). Distribution and quantile functions, ranks and signs in $\mathbb{R}^d$: a measure transportation approach. Working Papers ECARES  2017-34, ULB,  \url{https://ideas.repec.org/p/eca/wpaper/2013-258262.html}.\\

\noindent Hallin, M.,  del Barrio, E., Cuesta Albertos, J. and Matr\'an, C. (2019).  Center-outward distribution functions, quantiles, ranks, and signs in  $\mathbb{R}^d$. \url{https://arxiv.org/abs/1806.01238}. \\

\noindent Hill, B.M. (1975). A simple general approach to inference about the tail of a distribution.  {\it Annals of Statistics},  3, 1163--1174.\\

\noindent Kim, M. and Lee, S. (2017). Estimation of the tail exponent of multivariate regular variation. {\it Annals of the Institute of  Statistical Mathematics}, 69, 945--968.\\

\noindent Kusuoka, S. (2001). On law invariant coherent risk measures. {\it Advances in Mathematical Economics}, 3, 83--95.\\

\noindent Li, R., Fang, K.T., and Zhu, L.X. (1997). Some QQ probability plots to test spherical and elliptical symmetry.
\textit{Journal of Computational and Graphical Statistics}, {6}, 435--450.\\

\noindent  McCann, R. J. (1995). Existence and uniqueness of monotone measure-preserving maps. \textit{Duke Math. Journal},  {80}, 309--323.}\\

\noindent Rockafellar, R.T. (1966). Characterization of the subdifferential of convex functions. {\it Pacific Journal of Mathematics}, 17, 497--510.\\ 

\noindent Rockafellar, R.T. (1970). {\it Convex Analysis}. Princeton University Press.\\ 

\noindent R\"uschendorf, L. (2006). Law invariant convex risk measures for portfolio vectors. {it Statistics \& Decisions}, 24, 97--108.\\

\noindent Villani, C. (2009). {\it Optimal Transport: Old and New.} Grundlehren der Mathematischen Wissenschaften, Springer-Verlag, Heidelberg. \bigskip\bigskip\bigskip

\noindent

\end{document}